\documentclass[10pt]{amsart}
\usepackage{amsfonts,amssymb,amscd,amsmath,enumerate,verbatim,calc}
\usepackage{mathrsfs}
\numberwithin{equation}{section}
\newtheorem{theorem}{Theorem}[section]
\newtheorem{corollary}[theorem]{Corollary}

\newtheorem{proposition}[theorem]{Proposition}
\newtheorem{definition}[theorem]{Definition}

\newtheorem{remark}[theorem]{Remark}

\begin{document}

	\title[Statistical conformal Killing Vector Fields for FLRW Space-Time]
	{Statistical conformal Killing Vector Fields for FLRW Space-Time}

	\bibliographystyle{amsplain}
	
	\author[Esmaeil Peyghan, Leila Nourmohammadifar and Damianos Iosifidis]{ E. Peyghan, L. Nourmohammadifar  and D. Iosifidis}
	\address{Department of Mathematics, Faculty of Science, Arak University,
		Arak, 38156-8-8349, Iran.}
	\email{e-peyghan@araku.ac.ir,\ l.nourmohammadi@gmail.com}
	\address{Laboratory of Theoretical Physics, Institute of Physics, University of Tartu, W. Ostwaldi 1, 50411 Tartu, Estonia.}
	\email{damianos.iosifidis@ut.ee}

	



	\begin{abstract}
	The classification of conformal Killing vector fields for FLRW space-time from Riemannian point of view was done by Maartens-Maharaj in \cite{Maartens1986}. In this paper, we introduce conformal Killing vector fields from a new point of view for the FLRW space-time. 
In particular, we consider three cases for the conformal factor. Then, it is shown that there exist nine conformal vector fields on FLRW in total, such that six of them are Killing and the rest being non-Killing conformal vector fields. 	Consequently, by recalling the concept of statistical conformal  Killing vector fields introduced in \cite{SP}, we classify statistical structures with repsect to which these vector fields are conformal  Killing.
We also obtain the form of affine connections that feature a vanishing Lie derivative with respect to these conformal Killing vector fields.
Imposing the torsion-free and the Codazzi conditions on these connections,  we study  statistical structures on FLRW.  
Finally, for torsionful connections   we study the vanishing of the Lie derivative of the torsion tensor with respect to these conformal Killing vector fields and derive the conditions under which this is valid.
		
	\end{abstract}

	\maketitle

	\section{Introduction}
	
	Killing vector fields are of prominent importance in General Theory of Relativity (GR) and its extensions since they are associated with the existence of conserved currents. More specifically, if $X$ is a Killing vector field (i.e. $\pounds_Xg=0$) and $u$ the particle's 4-velocity then the quantity $X_{i}u^{i}$ is conserved along the particle's trajectory. Conformal Killing Vector Fields (CKVF), namely vector fields with the property $\pounds_Xg \propto g$ are also quite important in Physics. For instance, Hall and Capocci studied conformal (including homothetic and Killing) vector fields on a three-dimensional space–time \cite{hall}. Moreover, they explored the values of the dimensions of the Lie algebras of conformal, homothetic, and Killing vector fields under less restrictive conditions (the technique used was based on a similar one in general relativity \cite{Hall1,Hall:1991wy}). The conformal structure preserves the angles between vectors and consequently leaves the light cone intact. CKVF's have a wide range of applications from  GR Conformal Field Theories (CFTs) and also Holography (see for instance \cite{Ginsparg:1988ui}). These vector fields for some well-known space–times have been studied (see \cite{KHBK, KHBK1, MM}, for instance).
	
	On large scales our Universe is spatially isotropic and homogeneous and as such it is described by the so-called Friedmann-Lemaitre-Robertson-Walker (FLRW) metric \cite{Friedman:1922kd}. Using the form of the FLRW metric and the Einstein equations one then obtains the evolution equations for the scale factor (Friedmann equations) which describes how the Universe expands/contracts with the passing of time. Therefore, the FLRW space-time\footnote{We also have the Bianchi Cosmologies which are generalizations of FLRW space-time with less symmetries. For the study of CKVF's in such spaces see \cite{Qazi}.} is quite of importance in the study of Cosmology.  In this study we revisit the derivation of CKVF's in FLRW space-times. Let us note that the set of such vector fields was known in the literature \cite{Maartens1986}\footnote{See also the more recent \cite{Kehagias:2013xga} for the $k=0$ case.}. However, our approach here is different. We follow a more didactic step by step derivation of the CKVF's of the Robertson-Walker metric for all three cases $k=0, \pm 1$. Let us note that in this construction the maximally symmetric space is not the full 4-dim space-time but rather the spatial $t=const.$ 3-dim sub-manifold (see also \cite{HariDass:2015plk} for a relevant discussion). Consequently the scale factor (see FLRW metric below) is not constrained by the CKVF's  and is only obtained after solving the metric field equations at hand.

 Finally, we study Statistical CKVF's for the FLRW space-time. The motive for such an endeavor is supported by the nice properties that Statistical Manifolds exhibit. Indeed, the very fact that on such manifolds the inner products between vectors (and consequently their lengths) are left intact, provides a well motivated physical background to study gravity theories and their cosmology\footnote{In addition, it was recently shown \cite{Iosifidis:2023ftc}  that a bi-connection formulation of gravity sets itself as a statistical manifold. This ultimately suggests an intriguing correspondence between information geometry and gravity. }. Thus, even though the whole space has non-metricity (i.e. the covariant derivative of the metric does not vanish), the introduction of the dual connection $\nabla^{\star}$ succeeds in keeping the inner products unchanged when combined with the action of the initial connection $\nabla$.
Let us recall that the form of the Robertson-Walker metric  is derived by demanding spatial isotropy, namely a vanishing Lie derivative of the metric along the spatial directions. The same symmetry we demand when considering statistical manifolds but now we also extend it and demand isotropy for the connection coefficients as well. In the general case (with torsion) such a connection has 5 degrees of  freedom (see \cite{I}), and for the case of statistical manifolds this number is reduced accordingly.
  For such manifolds we find the form of the scale factor which is compatible with the Codazzi equations. Namely, we give the condition on the scale factor in order to admit a  statistical structure. Furthermore, we find the form of the scale factor for which the statistical structure is compatible with the 9 CKVF's of the FLRW space-time. Lastly, we study the symmetries of the torsion tensor and establish the conditions upon which its Lie derivative with respect to the CKVF's vanishes.

	\section{Preliminaries}
	In metric-affine gravitational theories  \cite{Hehl:1994ue}, the basic dynamical objects are the metric $g_{\mu\nu}$ and the connection $\Gamma^\kappa_{\mu\nu}$. The
	fundamental tensors that can be constructed with the help of these objects are the curvature $R^\kappa_{\lambda\mu\nu}$,  the torsion $T^\lambda_{\mu\nu}$, 
	and the non-metricity $Q_{\lambda\mu\nu}$, whose components are given respectively by
	
	\begin{align}
	&R^\kappa_{\lambda\mu\nu}=\partial_\lambda\Gamma^\kappa_{\mu\nu}-\partial_\mu\Gamma^\kappa_{\lambda\nu}+\Gamma^\sigma_{\mu\nu}\Gamma^\kappa_{\lambda\sigma}-\Gamma^\sigma_{\lambda\nu}\Gamma^\kappa_{\mu\sigma},\\
	&T^\lambda_{\mu\nu}=\Gamma^\lambda_{\mu\nu}-\Gamma^\lambda_{\nu\mu},\label{3999}\\
	&Q_{\lambda\mu\nu} =\nabla_\lambda g_{\mu\nu}= {\partial_\lambda g_{\mu\nu}} - \Gamma^{\sigma}_{\lambda\mu}g_{\sigma\nu} -\Gamma^{\sigma}_{\lambda\nu}g_{\mu\sigma}\label{3},
	\end{align}
	where $x^\lambda$ are the coordinate components upon the manifold, $\partial_\lambda=\frac{\partial}{\partial x^\lambda}$ and $\nabla_\lambda$ is used to denote the covariant derivative with respect to the affine connection $\Gamma^\kappa_{\lambda\nu}$. In the case of a symmetric (or torsion-free) connection the torsion tensor is zero,  i.e. $T^\lambda_{\mu\nu}=0$.
	This, together with the condition $Q_{\lambda\mu\nu}=0$, results in the well-known metric theories of gravity where the connection is uniquely determined and is none other than the Levi-Civita connection. When $Q_{\lambda\mu\nu}$ is totally symmetric (i.e., $Q_{\lambda\mu\nu}=Q_{\mu\lambda\nu}=Q_{\nu\lambda\mu}$), we have a Codazzi connection. A torsion-free and Codazzi connection is called a statistical connection. 
	
Modern cosmology is based on the Cosmological principle, whose symmetries call for the Friedmann-Lemaitre-Robertson-Walker (FLRW) metric \cite{M, W}, one of the most influential solutions to Einstein’s equations. FLRW is a special member of the class of spherically-symmetric space-times often used in problems of gravitational collapse or expansion.

The FLRW metric is conventionally written in the form
\begin{align}\label{1}
g=-N^2(t)dt^2+a^2(t)\left[\frac{dr^2}{\chi^2}+r^2(d\theta^2+sin^2\theta d\phi^2)\right],
\end{align}
where $\chi=\sqrt{1-\kappa r^2}$,  $a(t)$ is the scale factor and the spatial curvature is specified
by the constant $\kappa$, which takes the values $\kappa = 0, \pm 1$ for Euclidean, spherical and hyperbolic
geometries, respectively. In addition $N(t)$ is the lapse function which can be set to unity after a reparametrization of the time coordinate.

The Lie derivative of the metric $g$ with respect to a vector field $X$ is computed as follows
\begin{align}\label{1'}
(	\pounds_Xg)_{ij}=X^n\partial_ng_{ij}+g_{nj}\partial_jX^n+g_{ni}\partial_iX^n.
\end{align}
A vector field	$X$ is called a conformal vector field if $\pounds_Xg=\rho g$, where $\rho$ is a function on $M$, which is called the conformal factor. If $\rho$ is a constant function then $X$ is called  homothetic. In the special case where $\rho=0$, we say that $X$ is a Killing vector field. 
It is
called proper homothetic if $\rho\equiv$constant$\neq 0$  and proper conformal if $X$ is not homothetic \cite{hall}.
Recently, Samereh-Peyghan introduced the concept of conformal Killing vector fields on statistical manifolds (see \cite{SP}). They called a vector field on a statistical manifold, a conformal Killing vector field (CKVF) if $\pounds_Xg=\rho g$ and $\pounds_XK=0$, where $K=\nabla-\hat{\nabla}$ is the different tensor of a statistical connection $\nabla$ and the Levi-Civita connection $\hat{\nabla}$. Indeed, in this case, we have the CKVF's of the Riemmanian approach  together with the symmetry of $K$. It is necessary to mention that this definition can be considered from two different perspectives. In the first view, considering a fixed statistical structure $K$
on a manifold $(M,g)$, one can obtain the set of conformal vector fields for the  statistical manifold $(M, g, K)$. In the second point of view, first we find the vector fields that satisfy the condition $\pounds_Xg=\rho g$, and then we get statistical structures for which such vector fields  are conformal. In this paper, we consider the second point of view.
	\section{CKVF's of FLRW metric}
	In this section we study the  CKVF's  of the  FLRW metric. When $X$ is a CKVF, the conformal factor $\rho$ is a function with respect to $(r, \phi, \theta)$. As the study of CKVF's in this case is quite complicated, we focus on conformal factors with two variables only.  
\subsection{$\rho$ is a function with respect to $(r, \phi)$}
Let $X=X^1\partial_r+X^2\partial_\phi+X^3\partial_\theta$, where $X^1,X^2$ and $X^3$ are functions
 of the variables $r,\phi, \theta$. If $X$ is a CKVF with conformal factor $\rho(r,\phi)$, then from (\ref{1'}) we get the following system of PDE's:
\begin{align*}
&1)\ \frac{2\kappa r}{\chi^2}X^1+2\partial_rX^1=\rho(r,\phi),\\
&2)\  r^2\sin^2\theta\partial_r X^2+ \frac{1}{\chi^2}\partial_\phi X^1=0,\\
&3)\  r^2\partial_r X^3+\frac{1}{\chi^2} \partial_\theta X^1=0,\\
&4)\ 2X^1\sin\theta+2X^3r\cos\theta+2r\sin\theta\partial_\phi X^2=\rho(r,\phi) r\sin\theta,\\
&5)\ \partial_\phi X^3+\partial_\theta X^2\sin^2\theta=0,\\
&6)\ 2X^1+2r\partial_\theta X^3=\rho(r,\phi) r.
\end{align*}
Applying $\partial_\theta$ to both sides of (1), we have
\[
\frac{2\kappa r}{\chi^2}\partial_\theta X^1+2 \partial_r\partial_\theta X^1=0.
\]
The above differential equation has the following solution
\begin{align}\label{A01;}
X^1=\chi\int A(\theta,\phi)d\theta+B(r,\phi).
\end{align}
Setting the above relation in (1), we get
\begin{align}\label{A02;}
\frac{2\kappa r}{\chi^2}B(r,\phi)+2\partial_rB(r,\phi)=\rho(r, \phi).
\end{align} 
(\ref{A01;}) and (3) imply
\begin{align}\label{A01;;}
X^3=\frac{\chi}{r} A(\theta,\phi)+C(\theta,\phi).
\end{align}
Putting (\ref{A01;}) and (\ref{A01;;}) in (6), it follows
\[
\chi\int A(\theta,\phi)d\theta+B(r,\phi)+{\chi} \partial_\theta A(\theta,\phi)+r\partial_\theta C(\theta,\phi)=\frac{r}{2}\rho(r, \phi).
\]
Differentiating the above equation with respect to $\theta$, we obtain 
\begin{align}\label{L;}
\chi A(\theta,\phi)+{\chi} \partial^2_\theta A(\theta,\phi)+r\partial^2_\theta C(\theta,\phi)=0.
\end{align}
Applying $\partial^2_r$ to the above equation, we get
\begin{align}\label{L1;}
\partial^2_\theta A(\phi,\theta)+A(\phi,\theta)=0.
\end{align}
The above equation has the following solution: 
\begin{align}\label{80;}
A(\phi,\theta)=A_1(\phi)\cos\theta+A_2(\phi)\sin\theta.
\end{align}
From (\ref{L;}) and (\ref{L1;}), we have
\begin{align}\label{800;}
C(\phi,\theta)=\theta C_1(\phi)+C_2(\phi).
\end{align}
Substituting (\ref{80;}) and (\ref{800;}) in (\ref{A01;}) and (\ref{A01;;}) gives us 
\begin{align}
X^1=&\chi(A_1(\phi)\sin\theta-A_2(\phi)\cos\theta)+B(r,\phi),\label{AA01;}\\
X^3=&\frac{\chi}{r} (A_1(\phi)\cos\theta+A_2(\phi)\sin\theta)+\theta C_1(\phi)+C_2(\phi).\label{AA03;}
\end{align}
On the other hand, (4) combined with  (6) yields
\begin{align}\label{AAA03;}
\partial_\theta X^3-\cot\theta X^3=\partial_\phi X^2.
\end{align} 
Putting (\ref{AA03;}) in (\ref{AAA03;}), we get
\[
-\frac{\chi}{r} A_1(\phi)\csc\theta+ C_1(\phi)-\theta\cot\theta  C_1(\phi)-\cot\theta C_2(\phi)=\partial_\phi X^2.
\]
Integrating the above equation with respect to $\phi$ implies
\begin{align}\label{L22;}
X^2=-\frac{\chi}{r} \csc\theta \int A_1(\phi)d\phi+ (1-\theta\cot\theta)\int  C_1(\phi)d\phi-\cot\theta \int C_2(\phi)d\phi+D(r,\theta).
\end{align}
Plugging (\ref{AA01;}) and (\ref{L22;}) in (2), we have
\begin{align*}
&	\sin\theta\int A_1(\phi)d\phi+r^2{\chi}\sin^2\theta\partial_r D(r,\theta)+\frac{dA_1(\phi)}{d\phi}\sin\theta\\
&-\frac{dA_2(\phi)}{d\phi}\cos\theta+\frac{1}{{\chi}}\partial_\phi B(r,\phi)=0.
\end{align*}
Differentiating the above equation with respect to $\phi$ gives us
\begin{align}\label{L4;}
&	 A_1(\phi)\sin\theta+\frac{d^2A_1(\phi)}{d\phi^2}\sin\theta
-\frac{d^2A_2(\phi)}{d\phi^2}\cos\theta+\frac{1}{{\chi}}\partial^2_\phi B(r,\phi)=0.
\end{align}
Further differentiation of the latter with respect to $\theta$ now, gives
\begin{align*}
&	 \cos\theta (A_1(\phi)+\frac{d^2A_1(\phi)}{d\phi^2})
+\frac{d^2A_2(\phi)}{d\phi^2}\sin\theta=0.
\end{align*}
From the last equation we deduce
\begin{align*}
&A_1(\phi)+\frac{d^2A_1(\phi)}{d\phi^2}=0,\ \ \ \ \ \frac{d^2A_2(\phi)}{d\phi^2}=0,
\end{align*}
which has the following solutions:
\begin{align}\label{800,;}
A_1(\phi)=A_1\cos\phi+A_2\sin\phi, 
\end{align}
and 
\begin{align}\label{8000,;}
A_2(\phi)=a_1\phi+a_2,
\end{align}
where $A_1, A_2, a_1$ and $ a_2$ are constants.
Substituting (\ref{800,;}) and (\ref{8000,;}) in (\ref{L4;}), we obtain
\begin{align}\label{L4;;}
&	\partial^2_\phi B(r,\phi)=0,
\end{align}
that has the following solution:
\begin{align}\label{B;}
B(r,\phi)=B_1(r)\phi+B_2(r).
\end{align}
Setting (\ref{B;}) in (\ref{A02;}), it follows that
\begin{align*}
\frac{2\kappa r}{\chi^2}(B_1(r)\phi+B_2(r))+2\frac{dB_1(r)}{dr}\phi+2\frac{dB_2(r)}{dr}=\rho(r, \phi).
\end{align*} 
Applying $\partial^2_\phi$ to both sides of the above equation, we deduce
\begin{align}\label{L7;}
\rho(r, \phi)=\rho_1(r)\phi+\rho_2(r).
\end{align} 
From the above two equations we have 
\begin{align*}
\frac{dB_1(r)}{dr}+\frac{\kappa r}{\chi^2}B_1(r)=\frac{1}{2}\rho_1(r),
\end{align*}
and
\begin{align*}
\frac{dB_2(r)}{dr}+\frac{\kappa r}{\chi^2}B_2(r)=\frac{1}{2}\rho_2(r),
\end{align*}
which have the following solution:
\begin{align*}
{B_1(r)}=\chi\Big[\int\frac{1}{2\chi}\rho_1(r)dr+B_1 \Big],
\end{align*}
and
\begin{align*}
{B_2(r)}=\chi\Big[\int\frac{1}{2\chi}\rho_2(r)dr+B_2\Big].
\end{align*}
Putting (\ref{800,;})  and (\ref{B;}) in (\ref{AA01;}), (\ref{AA03;}) and (\ref{L22;}), we obtain
\begin{align}
X^1=&\chi\Big[(A_1\cos\phi+A_2\sin\phi)\sin\theta-(a_1\phi+a_2)\cos\theta\Big]\label{AA001;}\\
&+\chi\Big[\int\frac{1}{2\chi}\rho_1(r)dr+B_1\Big]\phi+\chi\Big[\int\frac{1}{2\chi}\rho_2(r)dr+B_2\Big],\nonumber\\
X^2=&-\frac{\chi}{r} \csc\theta (A_1\sin\phi-A_2\cos\phi)+ (1-\theta\cot\theta)\int  C_1(\phi)d\phi\label{AA002;}\\
&-\cot\theta \int C_2(\phi)d\phi+D(r,\theta),\nonumber\\
X^3=&\frac{\chi}{r} [(A_1\cos\phi+A_2\sin\phi)\cos\theta+(a_1\phi+a_2)\sin\theta]+\theta C_1(\phi)+C_2(\phi).\label{AA003;}
\end{align}
Setting (\ref{AA002;}) and (\ref{AA003;}) in (5), we get
\begin{align}\label{phi1}
&\frac{\chi}{r} a_1\sin\theta+\theta \frac{dC_1(\phi)}{d\phi}+\frac{dC_2(\phi)}{d\phi}+(-\cos\theta\sin\theta+\theta)\int  C_1(\phi)d\phi\\
&+\int C_2(\phi)d\phi+\sin^2\theta \partial_\theta D(r,\theta)=0.\nonumber
\end{align}
Differentiating the above equation with respect to $\phi$ implies
\begin{align*}
&\theta \frac{d^2C_1(\phi)}{d\phi^2}+\frac{d^2C_2(\phi)}{d\phi^2}+(-\cos\theta\sin\theta+\theta)  C_1(\phi)+ C_2(\phi)=0.
\end{align*}
Applying $\partial^2_\theta$ to the above equation, we get
\begin{align}\label{phi2}
C_1(\phi)=0.
\end{align}
From the above two equations we obtain
\begin{align*}
& \frac{d^2C_2(\phi)}{d\phi^2}+ C_2(\phi)=0,
\end{align*}
which has the following solution:
\begin{align}\label{phi3}
&C_2(\phi)=C_1\cos\phi+C_2\sin\phi.
\end{align}
Putting (\ref{phi2}) and (\ref{phi3}) in (\ref{phi1}), it follows that
\begin{align*}
&\frac{\chi}{r} a_1+\sin\theta \partial_\theta D(r,\theta)=0.
\end{align*}
The above equation yields
\begin{align}
D(r,\theta)=-\frac{\chi}{r} \ln(\csc\theta-\cot\theta) a_1+ D_1(r).
\end{align}
Using (\ref{AA001;}), (\ref{AA003;}), (\ref{phi2}) and (\ref{phi3}) in (6), we deduce
\begin{align}\label{N52'}
&\Big[\int\frac{1}{2\chi}\rho_1(r)dr+B_1\Big]\phi+\int\frac{1}{2\chi}\rho_2(r)dr+B_2=\frac{r}{2\chi}(\rho_1(r)\phi+\rho_2(r))r.
\end{align}
Differentiation of the above equation with respect to $\phi$ gives
\begin{align*}
\int\frac{1}{2\chi}\rho_1(r)dr+B_1=\frac{r}{2\chi}\rho_1(r)r.
\end{align*}
Taking the derivative of the latter  with respect to $r$, we have
\begin{align}
\frac{d\rho_1(r)}{dr}+\frac{kr}{{\chi^2}}\rho_1(r)=0,
\end{align}
which has the following solution:
\begin{align}\label{N53'}
\rho_1(r)=\mathcal{B}_1\chi,
\end{align}
where $\mathcal{B}_1$ is constant. (\ref{N52'}) and (\ref{N53'}) together imply 
\begin{align}\label{N53''}
\rho_2(r)=\mathcal{B}_2\chi,
\end{align}
where $\mathcal{B}_2$ is constant. Putting (\ref{N53'}) and (\ref{N53''}) in (\ref{N52'}), we deduce that $B_1=B_2=0$.
Setting the above equations in (\ref{AA001;})-(\ref{AA003;}), it follows
\begin{align}
X^1=&\chi[(A_1\cos\phi+A_2\sin\phi)\sin\theta-(a_1\phi+a_2)\cos\theta+\frac{1}{2}\mathcal{B}_1r\phi+\frac{1}{2}\mathcal{B}_2r]\label{AAA001;},\\
X^2=&-\frac{\chi}{r} \csc\theta (A_1\sin\phi-A_2\cos\phi)-\cot\theta(C_1\sin\phi-C_2\cos\phi)\label{AAA002;}\\
&-\frac{\chi}{r} \ln(\csc\theta-\cot\theta) a_1+ D_1(r),\nonumber\\
X^3=&\frac{\chi}{r} [(A_1\cos\phi+A_2\sin\phi)\cos\theta+(a_1\phi+a_2)\sin\theta]+C_1\cos\phi+C_2\sin\phi.\label{AAA003;}
\end{align}
Putting (\ref{AAA001;}) and (\ref{AAA002;}) in (2), we deduce
\[
\frac{1}{\chi}\big(\sin^2\theta\ln(\csc\theta-\cot\theta) -\cos\theta)a_1+r^2\sin^2\theta\frac{dD_1(r)}{dr}+\frac{r}{2\chi}\mathcal{B}_1=0.
\]
Differentiating the above equation with respect to $\theta$ implies that
\[
\frac{1}{\chi}\big(\sin2\theta\ln(\csc\theta-\cot\theta) +2\sin\theta)a_1+r^2\sin2\theta\frac{dD_1(r)}{dr}=0.
\]
Considering $\theta=\frac{\pi}{2}$ in the above equation, we get $a_1=0$ and consequently, $D_1(r)=D_1$, where $D_1$ is constant. So, the above two equations imply $\mathcal{B}_1=0$.
Hence (\ref{AAA001;})-(\ref{AAA003;}) reduce to
\begin{align}
X^1=&\chi[(A_1\cos\phi+A_2\sin\phi)\sin\theta-a_2\cos\theta+\frac{1}{2}\mathcal{B}_2r]\label{AAAA001;},\\
X^2=&-\frac{\chi}{r} \csc\theta (A_1\sin\phi-A_2\cos\phi)-\cot\theta(C_1\sin\phi-C_2\cos\phi)+ D_1,\label{AAAA002;}\\
X^3=&\frac{\chi}{r} [(A_1\cos\phi+A_2\sin\phi)\cos\theta+a_2\sin\theta]+C_1\cos\phi+C_2\sin\phi.\label{AAAA003;}
\end{align}
\begin{theorem}\label{A100}
The vector fields $X=X^1\partial_r+X^2\partial_\phi+X^3\partial_\theta$ are conformal vector fields on FLRW with a conformal factor $\rho(r, \phi)$ if and only if $X^i, i=1, 2, 3$ satisfy (\ref{AAAA001;})-(\ref{AAAA003;}) and $\rho$ is a function with respect to $r$, given by $\rho(r)=\mathcal{B}_2\chi$, where $\mathcal{B}_2$ is a constant. Moreover, when $\kappa=0$, the conformal vector field reduces to a homothetic vector field.
\end{theorem}
\subsection{$\rho$ is a function with respect to $(\phi, \theta)$}
Let $X=X^1\partial_r+X^2\partial_\phi+X^3\partial_\theta$ where $X^1,X^2$ and $X^3$ are functions
of the variables $r,\phi, \theta$. If $X$ is a conformal vector field with conformal factor $\rho(\phi, \theta)$, then  we get the following system
\begin{align*}
&1)\ \frac{2\kappa r}{\chi^2}X^1+2\partial_rX^1=\rho(\phi, \theta),\\
&2)\  r^2\sin^2\theta\partial_r X^2+ \frac{1}{\chi^2}\partial_\phi X^1=0,\\
&3)\  r^2\partial_r X^3+\frac{1}{\chi^2} \partial_\theta X^1=0,\\
&4)\ 2X^1\sin\theta+2X^3r\cos\theta+2r\sin\theta\partial_\phi X^2=\rho(\phi, \theta) r\sin\theta,\\
&5)\ \partial_\phi X^3+\partial_\theta X^2\sin^2\theta=0,\\
&6)\ 2X^1+2r\partial_\theta X^3=\rho(\phi, \theta) r.
\end{align*}
By integrating (1) with respect to $r$, we have
\begin{align}\label{L3'}
X^1=\chi\Big[\int\frac{1}{2\chi}\rho(\phi, \theta)dr+A(\phi, \theta)\Big].
\end{align}
{\bf Case 1. $\kappa\neq 0$}\\
In this case, (\ref{L3'}) has the following solution:
\begin{align}\label{L3''}
X^1=\chi\Big[\frac{1}{2\sqrt{\kappa }}\rho(\phi, \theta)\tan^{-1}(\frac{r\sqrt{\kappa }}{\chi})+A(\phi, \theta)\Big].
\end{align}
Applying (\ref{L3''}) in (2), it follows
\begin{align*}
\partial_r X^2=-\frac{1}{r^2\chi} \csc^2\theta\Big[\frac{1}{2\sqrt{\kappa }}\partial_\phi\rho(\phi, \theta)\tan^{-1}(\frac{r\sqrt{\kappa }}{\chi})+\partial_\phi A(\phi, \theta)\Big].
\end{align*}
Integrating the above equation with respect to $r$, we have
\begin{align}\label{L4'}
X^2=- \csc^2\theta\Big[\frac{1}{2\sqrt{\kappa }}\partial_\phi\rho(\phi, \theta)\int \frac{1}{r^2\chi}\tan^{-1}(\frac{r\sqrt{\kappa }}{\chi})dr-\frac{\chi}{r}\partial_\phi A(\phi, \theta)\Big]+B(\phi,\theta).
\end{align}
(\ref{L3''}) and (3) imply
\begin{align*}
\partial_r X^3=-\frac{1}{r^2\chi} \Big[\frac{1}{2\sqrt{\kappa }}\partial_\theta\rho(\phi, \theta)\tan^{-1}(\frac{r\sqrt{\kappa }}{\chi})+\partial_\theta A(\phi, \theta)\Big].
\end{align*}
Hence we have 
\begin{align}\label{L5'}
X^3=- \frac{1}{2\sqrt{\kappa }}\partial_\theta\rho(\phi, \theta)\int \frac{1}{r^2\chi}\tan^{-1}(\frac{r\sqrt{\kappa }}{\chi})dr+\frac{\chi}{r}\partial_\theta A(\phi, \theta)]+C(\phi,\theta).
\end{align}
Setting (\ref{L4'}) and (\ref{L5'}) in (5), it follows 
\begin{align*}
&- \frac{1}{\sqrt{\kappa }}\partial_\phi\partial_\theta\rho(\phi, \theta)\int \frac{1}{r^2\chi}\tan^{-1}(\frac{r\sqrt{\kappa }}{\chi})dr+2\frac{\chi}{r}\partial_\phi\partial_\theta A(\phi, \theta)+\partial_\phi C(\phi,\theta)\\
&+2\cot\theta\Big[ \frac{1}{2\sqrt{\kappa }}\partial_\phi\rho(\phi, \theta)\int \frac{1}{r^2\chi}\tan^{-1}(\frac{r\sqrt{\kappa }}{\chi})dr-\frac{\chi}{r}\partial_\phi A(\phi, \theta)\Big]+\sin^2\theta\partial_\theta B(\phi,\theta)=0.
\end{align*}
Differentiating the last relation with respect to $r$, we get
\begin{align*}
&- \frac{1}{\sqrt{\kappa }}\partial_\phi\partial_\theta\rho(\phi, \theta) \frac{1}{r^2\chi}\tan^{-1}(\frac{r\sqrt{\kappa }}{\chi})-\frac{2}{r^2\chi}\partial_\phi\partial_\theta A(\phi, \theta)\\
&+2\cot\theta[ \frac{1}{2\sqrt{\kappa }}\partial_\phi\rho(\phi, \theta)\frac{1}{r^2\chi}\tan^{-1}(\frac{r\sqrt{\kappa }}{\chi})+\frac{1}{r^2\chi}\partial_\phi A(\phi, \theta)]=0,
\end{align*}
thus we have
\begin{align}\label{L8'}
&- \frac{1}{\sqrt{\kappa }}\partial_\phi\partial_\theta\rho(\phi, \theta) \tan^{-1}(\frac{r\sqrt{\kappa }}{\chi})-2\partial_\phi\partial_\theta A(\phi, \theta)\\
&+2\cot\theta\Big[ \frac{1}{2\sqrt{\kappa }}\partial_\phi\rho(\phi, \theta)\tan^{-1}(\frac{r\sqrt{\kappa }}{\chi})+\partial_\phi A(\phi, \theta)\Big]=0\nonumber,
\end{align}
Differentiating the above equation with respect to $r$ yields
\begin{align*}
&\partial_\phi\partial_\theta\rho(\phi, \theta) 
-\cot\theta\partial_\phi\rho(\phi, \theta)=0,
\end{align*}
which has the solution: 
\begin{align}\label{L7''}
&\rho(\phi, \theta) 
=\sin\theta\int \rho_1(\phi)d\phi+\rho_2(\theta).
\end{align}
Putting (\ref{L7''}) in  (\ref{L8'}) we get 
\begin{align}\label{L7'}
&A(\phi, \theta) 
=\sin\theta\int A_1(\phi)d\phi+A_2(\theta).
\end{align}
Replacing (\ref{L7''}) and (\ref{L7'}) in (\ref{L3''}), (\ref{L4'}) and (\ref{L5'}), we have 
\begin{align}
X^1=&\chi\Big[\frac{1}{2\sqrt{\kappa }}(\sin\theta\int \rho_1(\phi)d\phi+\rho_2(\theta))\tan^{-1}(\frac{r\sqrt{\kappa }}{\chi})+\sin\theta\int A_1(\phi)d\phi+A_2(\theta)\Big],\label{AphiL7'}\\
X^2=&- \frac{1}{2\sqrt{\kappa }}\rho_1(\phi)\csc\theta\int \frac{1}{r^2\chi}\tan^{-1}(\frac{r\sqrt{\kappa }}{\chi})dr+\frac{\chi}{r}\csc\theta A_1(\phi)+B(\phi,\theta),\label{BphiL7'}\\
X^3=&- \frac{1}{2\sqrt{\kappa }}\Big(\cos\theta\int \rho_1(\phi)d\phi+\frac{d\rho_2(\theta)}{d\theta}\Big)\int \frac{1}{r^2\chi}\tan^{-1}(\frac{r\sqrt{\kappa }}{\chi})dr\label{CphiL7'}\\
&+\frac{\chi}{r}\Big(\cos\theta\int A_1(\phi)d\phi+\frac{dA_2(\theta)}{d\theta})+C(\phi,\theta\Big)\nonumber.
\end{align}
Also, (4) and (6) imply
\begin{align}\label{phi30}
\partial_\theta X^3-\cot\theta X^3=\partial_\phi X^2.
\end{align}
Substituting (\ref{BphiL7'}) and (\ref{CphiL7'}) in (\ref{phi30}), we obtain 
\begin{align}\label{L71}
& \frac{1}{2\sqrt{\kappa }}\Big(\csc\theta\int \rho_1(\phi)d\phi-\frac{d^2\rho_2(\theta)}{d\theta^2}+\cot\theta\frac{d\rho_2(\theta)}{d\theta}+\csc\theta\frac{d\rho_1(\phi)}{d\phi}\Big)\int \frac{1}{r^2\chi}\tan^{-1}(\frac{r\sqrt{\kappa }}{\chi})dr\\
&-\csc\theta\frac{\chi}{r}\Big(\frac{d A_1(\phi)}{d\phi}+\int A_1(\phi)d\phi\Big)+\frac{\chi}{r}\Big(\frac{d^2A_2(\theta)}{d\theta^2}-\cot\theta\frac{dA_2(\theta)}{d\theta}\Big)+\frac{\partial C(\phi,\theta)}{\partial\theta}\nonumber\\
&-\cot\theta C(\phi,\theta)-\frac{\partial B(\phi,\theta)}{\partial\phi}=0.\nonumber
\end{align}
Differentiating the above relation with respect to $r$ gives
\begin{align*}
& \frac{1}{2\sqrt{\kappa }}\Big(\csc\theta\int \rho_1(\phi)d\phi-\frac{d^2\rho_2(\theta)}{d\theta^2}+\cot\theta\frac{d\rho_2(\theta)}{d\theta}+\csc\theta\frac{d\rho_1(\phi)}{d\phi}\Big) \frac{1}{r^2\chi}\tan^{-1}(\frac{r\sqrt{\kappa }}{\chi})\\
&+\csc\theta\frac{1}{r^2\chi}\Big(\frac{d A_1(\phi)}{d\phi}+\int A_1(\phi)d\phi\Big)-\frac{1}{r^2\chi}(\frac{d^2A_2(\theta)}{d\theta^2}-\cot\theta\frac{dA_2(\theta)}{d\theta})=0,
\end{align*}
which implies 
\begin{align}
& \frac{1}{2\sqrt{\kappa }}\Big(\csc\theta\int \rho_1(\phi)d\phi-\frac{d^2\rho_2(\theta)}{d\theta^2}+\cot\theta\frac{d\rho_2(\theta)}{d\theta}+\csc\theta\frac{d\rho_1(\phi)}{d\phi}\Big)\tan^{-1}(\frac{r\sqrt{\kappa }}{\chi})\label{phi33}\\
&+\csc\theta(\frac{d A_1(\phi)}{d\phi}+\int A_1(\phi)d\phi)-\frac{d^2A_2(\theta)}{d\theta^2}+\cot\theta\frac{dA_2(\theta)}{d\theta}=0.\nonumber
\end{align}
Applying $\partial_\phi\partial_r$ to both sides of the above equation, we obtain
\[
\rho_1(\phi)+\frac{d^2\rho_1(\phi)}{d\phi^2}=0,
\]
which has the following solution:
\begin{align}\label{phi32}
\rho_1(\phi)=\rho_3\cos\phi+\rho_4\sin\phi.
\end{align}
Setting (\ref{phi32}) in (\ref{phi33}), we deduce
\[
\cot\theta\frac{d\rho_2(\theta)}{d\theta}-\frac{d^2\rho_2(\theta)}{d\theta^2}=0.
\]
Hence we get
\begin{align}\label{phi34}
\rho_2(\theta)=-\rho_5\cos\theta+\rho_6.
\end{align}
Replacing (\ref{phi32}) and (\ref{phi34}) in  (\ref{phi33}) gives
\begin{align}\label{phi38}
\csc\theta\Big(\frac{d A_1(\phi)}{d\phi}+\int A_1(\phi)d\phi\Big)-\frac{d^2A_2(\theta)}{d\theta^2}+\cot\theta\frac{dA_2(\theta)}{d\theta}=0.
\end{align}
Differentiating the last relation with respect to $\phi$, we get
\[
\frac{d^2 A_1(\phi)}{d\phi^2}+ A_1(\phi)=0,
\]
which results in
\begin{align}\label{phi37}
A_1(\phi)=A_3\cos\phi+A_4\sin\phi.
\end{align}
Using (\ref{phi37}) in (\ref{phi38}), it follows that
\begin{align}\label{phi39}
A_2(\theta)=-A_5\cos\theta+A_6.
\end{align}
Substituting (\ref{phi32}), (\ref{phi34}), (\ref{phi37}) and (\ref{phi39}) in (\ref{AphiL7'}), (\ref{BphiL7'}) and (\ref{CphiL7'}), we obtain
\begin{align}
X^1=&\chi\Big[\frac{1}{2\sqrt{\kappa }}(\sin\theta(\rho_3\sin\phi-\rho_4\cos\phi)-\rho_5\cos\theta+\rho_6)\tan^{-1}(\frac{r\sqrt{\kappa }}{\chi})\label{AAphiL7'}\\
&+\sin\theta(A_3\sin\phi-A_4\cos\phi)-A_5\cos\theta+A_6\Big],\nonumber\\
X^2=&- \frac{1}{2\sqrt{\kappa }}(\rho_3\cos\phi+\rho_4\sin\phi)\csc\theta\int \frac{1}{r^2\chi}\tan^{-1}(\frac{r\sqrt{\kappa }}{\chi})dr\label{BBphiL7'}\\
&+\frac{\chi}{r}\csc\theta (A_3\cos\phi+A_4\sin\phi)+B(\phi,\theta),\nonumber\\
X^3=&- \frac{1}{2\sqrt{\kappa }}\Big(\cos\theta(\rho_3\sin\phi-\rho_4\cos\phi)+\rho_5\sin\theta\Big)\int \frac{1}{r^2\chi}\tan^{-1}(\frac{r\sqrt{\kappa }}{\chi})dr\label{CCphiL7'}\\
&+\frac{\chi}{r}\Big(\cos\theta(A_3\sin\phi-A_4\cos\phi)+A_5\sin\theta\Big)+C(\phi,\theta)\nonumber.
\end{align}
Combining (\ref{AAphiL7'}), (\ref{CCphiL7'}) and (1), we obtain
\begin{align}\label{phi90}
&\frac{\chi}{r}\Big[\frac{1}{2\sqrt{\kappa }}(\sin\theta(\rho_3\sin\phi-\rho_4\cos\phi)-\rho_5\cos\theta+\rho_6)\tan^{-1}(\frac{r\sqrt{\kappa }}{\chi})+A_6\Big]\\
&
- \frac{1}{2\sqrt{\kappa }}(-\sin\theta(\rho_3\sin\phi-\rho_4\cos\phi)+\rho_5\cos\theta)\int \frac{1}{r^2\chi}\tan^{-1}(\frac{r\sqrt{\kappa }}{\chi})dr+\frac{\partial C(\phi,\theta)}{\partial\theta}\nonumber\\
&
=\frac{1}{2}\Big(\sin\theta(\rho_3\sin\phi-\rho_4\cos\phi)-\rho_5\cos\theta+\rho_6\Big).\nonumber
\end{align}
Differentiating the above equation with respect to $r$, we get
\begin{align*}
&\frac{1}{2\sqrt{\kappa }}\Big(-\frac{1}{r^2\chi}\tan^{-1}(\frac{r\sqrt{\kappa }}{\chi})+\frac{\sqrt{\kappa}}{r}\Big)\Big(\sin\theta(\rho_3\sin\phi-\rho_4\cos\phi)-\rho_5\cos\theta+\rho_6\Big)-\frac{1}{r^2\chi}A_6\\
&
- \frac{1}{2\sqrt{\kappa }}(-\sin\theta(\rho_3\sin\phi-\rho_4\cos\phi)+\rho_5\cos\theta) \frac{1}{r^2\chi}\tan^{-1}(\frac{r\sqrt{\kappa }}{\chi})=0,
\end{align*}
which gives 
\begin{align}\label{phi41}
&\frac{1}{2\sqrt{\kappa }}\big(-\frac{1}{r^2\chi}\tan^{-1}(\frac{r\sqrt{\kappa }}{\chi})+\frac{\sqrt{\kappa}}{r}\big)\rho_6+\frac{1}{2r}(\sin\theta(\rho_3\sin\phi-\rho_4\cos\phi)-\rho_5\cos\theta)\\
&-\frac{1}{r^2\chi}A_6=0.\nonumber
\end{align}
Applying $\frac{\partial}{\partial{\theta}}$ to  the above equation implies
\begin{align*}
\cos\theta(\rho_3\sin\phi-\rho_4\cos\phi)+\rho_5\sin\theta=0,
\end{align*}
which gives us 
\begin{align}\label{phi40}
\rho_3=\rho_4=\rho_5=0.
\end{align}
Setting (\ref{phi40}) in (\ref{phi41}), we deduce
\begin{align}\label{phi42}
A_6=\rho_6=0.
\end{align}
So, using the above equation in  (\ref{L7''}) gives
\begin{align}\label{L7'';}
\rho(\phi, \theta) 
=0.
\end{align}
On the other hand, (\ref{phi90}) implies
\[
\frac{\partial C(\phi,\theta)}{\partial\theta}=0.
\]
Hence we have 
\begin{align}\label{pl}
C(\phi,\theta)=C_1(\phi).
\end{align}
Also, (\ref{L71}) yields
\[
\cot\theta C_1(\phi)+\frac{\partial B(\phi,\theta)}{\partial\phi}=0,
\]
which gives 
\begin{align}\label{phi37799}
B(\phi,\theta)=-\cot\theta\int C_1(\phi)d\phi+B_1(\theta).
\end{align}
Moreover, substituting (\ref{BBphiL7'}) and (\ref{CCphiL7'}) in (5), one sees that
\begin{align}\label{phi3779}
\frac{dC_1(\phi)}{d\phi}+\int C_1(\phi)d\phi+\sin^2\theta\frac{dB_1(\theta)}{d\theta}=0.
\end{align}
Differentiating the above equation with respect to $\phi$, we deduce 
\[
\frac{d^2C_1(\phi)}{d\phi^2}+ C_1(\phi)=0,
\]
which gives us 
\begin{align}\label{phi377}
C_1(\phi)=C_1\cos\phi+C_2\sin\phi.
\end{align}
Plugging (\ref{phi377}) in (\ref{phi3779}), it follows that
\[
B_1(\theta)=B_1,
\]
where $B_1$ is a constant. Hence (\ref{phi37799}) gives
\begin{align}
B(\phi,\theta)=-\cot\theta(C_1\sin\phi-C_2\cos\phi)+B_1.
\end{align}
Substituting the above equations in (\ref{AAphiL7'}), (\ref{BBphiL7'}) and (\ref{CCphiL7'}), we end up with
\begin{align}
X^1=&\chi[\sin\theta(A_3\sin\phi-A_4\cos\phi)-A_5\cos\theta],\label{pphi377}\\
X^2=&\frac{\chi}{r}\csc\theta (A_3\cos\phi+A_4\sin\phi)-\cot\theta(C_1\sin\phi-C_2\cos\phi)+B_1,\\
X^3=&\frac{\chi}{r}(\cos\theta(A_3\sin\phi-A_4\cos\phi)+A_5\sin\theta)+C_1\cos\phi+C_2\sin\phi.\label{ppphi377}
\end{align}

{\bf Case 2. $\kappa= 0$}\\
In this case, (\ref{L3'}) has the following solution:
\begin{align}\label{L3}
X^1=\frac{1}{2}\rho(\phi, \theta)r+A(\phi, \theta).
\end{align}
Setting the above equation in (2) and (3), we obtain 
\begin{align}\label{L2}
X^2=-\csc^2\theta(\frac{1}{2}\partial_\phi\rho(\phi, \theta)\ln r-\frac{1}{r}\partial_\phi A(\phi, \theta))+C(\phi, \theta),
\end{align}
and
\begin{align}\label{L1}
X^3=-\frac{1}{2}\partial_\theta\rho(\phi, \theta)\ln r+\frac{1}{r}\partial_\theta A(\phi, \theta)+B(\phi, \theta).
\end{align}
Putting (\ref{L3}) and  (\ref{L1}) in (6), it follows that
\begin{align}\label{L7}
A(\phi, \theta)+r[-\frac{1}{2}\partial^2_\theta\rho(\phi, \theta)\ln r+\frac{1}{r}\partial^2_\theta A(\phi, \theta)+\partial_\theta B(\phi, \theta)]=0.
\end{align}
By differentiating twice the above equation with respect to
$r$ it follows that  $\partial_\theta B(\phi, \theta)=0$ and $\partial^2_\theta\rho(\phi, \theta)=0$. Hence we have 
\begin{align}\label{L5}
B(\phi, \theta)=B(\phi).
\end{align}
and 
\begin{align}\label{L6}
\rho(\phi, \theta)=\rho_1(\phi)\theta+\rho_2(\phi).
\end{align}
Setting (\ref{L5}) and  (\ref{L6}) in (\ref{L7}), we get 
\begin{align*}
\partial^2_\theta A(\phi, \theta)+A(\phi, \theta)=0.
\end{align*}
The above equation has the following solution: 
\begin{align*}
A(\phi, \theta)=A_1(\phi)\cos\theta+A_2(\phi)\sin\theta,
\end{align*}
where $A_1$ and $ A_2$ are constant. Using the above equations in (\ref{L3})-(\ref{L1}), we get
\begin{align}
X^1&=\frac{r}{2}(\rho_1(\phi)\theta+\rho_2(\phi))+A_1(\phi)\cos\theta+A_2(\phi)\sin\theta,\label{L0i}\\
X^2&=-\csc^2\theta[\frac{1}{2}(\frac{d\rho_1(\phi)}{d\phi}\theta+\frac{d\rho_2(\phi)}{d\phi})\ln r-\frac{1}{r} (\frac{dA_1(\phi)}{d\phi}\cos\theta+\frac{dA_2(\phi)}{d\phi} \sin\theta)]+C(\phi, \theta),\label{L0ii}\\
X^3&=-\frac{1}{2}\rho_1(\phi)\ln r+\frac{1}{r}(-A_1(\phi)\sin\theta+A_2(\phi)\cos\theta)+B(\phi).\label{L0iii}
\end{align}
Putting (\ref{L0ii}) and (\ref{L0iii}) in (5) gives us
\begin{align}\label{L11}
&-\frac{d\rho_1(\phi)}{d\phi}\ln r-\frac{2}{r} \frac{dA_1(\phi)}{d\phi}\sin\theta+\frac{dB(\phi)}{d\phi}+2\cot\theta[\frac{1}{2}(\frac{d\rho_1(\phi)}{d\phi}\theta+\frac{d\rho_2(\phi)}{d\phi})\ln r-\frac{1}{r} \frac{dA_1(\phi)}{d\phi}\cos\theta]\\
&+\sin^2\theta\partial_\theta{C(\phi, \theta)}=0.\nonumber
\end{align}
Applying $\frac{d^2}{d r^2}$ to the above equation we have
\begin{align}\label{L9}
&-\frac{d\rho_1(\phi)}{d\phi}+\cot\theta(\frac{d\rho_1(\phi)}{d\phi}\theta+\frac{d\rho_2(\phi)}{d\phi})=0.
\end{align}
Setting $\theta=\frac{\pi}{2}$ in the latter, we get the differential equation 
\begin{align*}
\frac{d\rho_1(\phi)}{d\phi}=0,
\end{align*}
which has the solution 
\begin{align}\label{L8}
\rho_1(\phi)=\rho_1,
\end{align}
where $\rho_1$ is constant. Setting (\ref{L8}) in (\ref{L9}) we get 
$
\frac{d\rho_2(\phi)}{d\phi}=0,
$ and consecuently, 
\begin{align}\label{L10}
\rho_2(\phi)=\rho_2,
\end{align}
where $\rho_2$ is constant.
Plugging (\ref{L8}) and  (\ref{L10})  in (\ref{L11}) we deduce
\begin{align}\label{L14}
&-{2} \frac{dA_1(\phi)}{d\phi}(\sin\theta+\cot\theta\cos\theta)+r\frac{dB(\phi)}{d\phi}
+r\sin^2\theta\partial_\theta{C(\phi, \theta)}=0.
\end{align}
Differentiation of the above equation with respect to $r$ implies 
\begin{align}\label{L12}
&\frac{dB(\phi)}{d\phi}
+\sin^2\theta\partial_\theta{C(\phi, \theta)}=0.
\end{align}
Then differentiating this with respect to $\theta$ gives
\begin{align*}
\sin\theta\partial^2_\theta{C(\phi, \theta)}+2\cos\theta\partial_\theta{C(\phi, \theta)}=0,
\end{align*}
which has the following solution:
\begin{align*}
{C(\phi, \theta)}=-{C_1(\phi)}\cot\theta+C_2(\phi).
\end{align*}
Substituting the last equation in (\ref{L12}), it follows that
\begin{align}\label{N22}
{C(\phi, \theta)}=\frac{dB(\phi)}{d\phi}\cot\theta+C_2(\phi).
\end{align}
(\ref{L14}) and  (\ref{L12})  yield $\frac{dA_1(\phi)}{d\phi}=0$. Hence we have 
\begin{align}\label{P260}
A_{1}(\phi)=A_1,
\end{align}
where ${A}_1$  is a constant. 
Putting  (\ref{L8}), (\ref{L10}), (\ref{N22}) and (\ref{P260}) in (\ref{L0i})-(\ref{L0iii}) we get
\begin{align}
X^1&=\frac{r}{2}(\rho_1\theta+\rho_2)+A_1\cos\theta+A_2(\phi)\sin\theta,\label{L00i}\\
X^2&=\frac{1}{r}\csc\theta\frac{dA_2(\phi)}{d\phi}+\frac{dB(\phi)}{d\phi}\cot\theta+C_2(\phi),\label{L00ii}\\
X^3&=-\frac{1}{2}\rho_1\ln r+\frac{1}{r}(-A_1\sin\theta+A_2(\phi)\cos\theta)+B(\phi).\label{L00iii}
\end{align}
Setting (\ref{L6}) and  (\ref{L00i})-(\ref{L00iii})   in (3) we obtain
\begin{align}\label{L15'p}
&2A_2(\phi)-\rho_1r\ln r\cos\theta+2rB(\phi)\cos\theta+2\frac{d^2A_2(\phi)}{d\phi^2}+2r\sin\theta\frac{dC_2(\phi)}{d\phi}+2r\cos\theta\frac{d^2B(\phi)}{d\phi^2}=0.
\end{align}
Differentiating the last equation with respect to $r$  gives
\begin{align}\label{L15}
&-\rho_1\ln r\cos\theta-\rho_1\cos\theta+2B(\phi)\cos\theta+2\sin\theta\frac{dC_2(\phi)}{d\phi}+2\cos\theta\frac{d^2B(\phi)}{d\phi^2}=0.
\end{align}
Applying $\frac{d}{dr}$ to  the above equation implies $\rho_1=0$. So (\ref{L6}) yields
\begin{align*}
{\rho(\phi, \theta)}=\rho_2.
\end{align*}
Hence (\ref{L15}) implies
\begin{align}\label{L15'}
&B(\phi)\cot\theta+\frac{dC_2(\phi)}{d\phi}+\cot\theta\frac{d^2B(\phi)}{d\phi^2}=0.
\end{align}
Differentiating the above with respect to $\theta$  we have
\[
B(\phi)+\frac{d^2B(\phi)}{d\phi^2}=0,
\]
which has the following solution:
\[
B(\phi)=B_1\cos\phi+B_2\sin\phi.
\]
Setting the last equation (\ref{L15'}), we get $C_2(\phi)=C_2$, where $C_2$ is constant. Therefore, (\ref{L15'p}), implies 
\[
A_2(\phi)=\mathcal{A}_1\cos\phi+\mathcal{A}_2\sin\phi,
\]
where $\mathcal{A}_1$ and $\mathcal{A}_2$ are constant. Substituting the above equations in (\ref{L00i})-(\ref{L00iii}), we finally deduce
\begin{align}
X^1&=\frac{r}{2}\rho_2+A_1\cos\theta+(\mathcal{A}_1\cos\phi+\mathcal{A}_2\sin\phi)\sin\theta,\label{qL00i}\\
X^2&=\frac{1}{r}\csc\theta(-\mathcal{A}_1\sin\phi+\mathcal{A}_2\cos\phi)+(-B_1\sin\phi+B_2\cos\phi)\cot\theta+C_2,\label{qL00ii}\\
X^3&=\frac{1}{r}(-A_1\sin\theta+(\mathcal{A}_1\cos\phi+\mathcal{A}_2\sin\phi)\cos\theta)+B_1\cos\phi+B_2\sin\phi.\label{qL00iii}
\end{align}
which immediately leads to the following result.
\begin{theorem}\label{A101}
	There exists no  proper conformal Killing vector field on FLRW space with
	 conformal factor of the form $\rho( \phi, \theta)$  (namely a function of  $ \phi$ and $ \theta$ only). More precisely, we have\\
	 i) If $\kappa\neq 0$, 
	the conformal vector fields $X=X^1\partial_r+X^2\partial_\phi+X^3\partial_\theta$ reduce to the Killing vector fields with components  $X^i, i=1, 2, 3$ satisfying (\ref{pphi377})-(\ref{ppphi377}).\\
		ii) If $\kappa=0$, 
	the conformal vector fields $X=X^1\partial_r+X^2\partial_\phi+X^3\partial_\theta$ reduce to the homothetic vector fields $X^i, i=1, 2, 3$ satisfying (\ref{qL00i})-(\ref{qL00iii}).
	 \end{theorem}
\subsection{$\rho$ is a function with respect to $(r, \theta)$ }
Let $X=X^1\partial_r+X^2\partial_\phi+X^3\partial_\theta$ where $X^1,X^2$ and $X^3$ are functions
in terms of the variables $r,\phi, \theta$. If $X$ is a conformal vector field with conformal factor $\rho(r, \theta)$, then  we get
\begin{align*}
&1)\ \frac{2\kappa r}{\chi^2}X^1+2\partial_rX^1=\rho(r, \theta),\\
&2)\  r^2\sin^2\theta\partial_r X^2+ \frac{1}{\chi^2}\partial_\phi X^1=0,\\
&3)\  r^2\partial_r X^3+\frac{1}{\chi^2} \partial_\theta X^1=0,\\
&4)\ 2X^1\sin\theta+2X^3r\cos\theta+2r\sin\theta\partial_\phi X^2=\rho(r, \theta) r\sin\theta,\\
&5)\ \partial_\phi X^3+\partial_\theta X^2\sin^2\theta=0,\\
&6)\ 2X^1+2r\partial_\theta X^3=\rho(r, \theta) r.
\end{align*}
Using (1), we get
\begin{align}\label{rA01;}
X^1=\chi\Big[\int \frac{1}{2\chi}\rho(r,\theta)dr+A(\phi,\theta)\Big].
\end{align}
Differentiating the above equation with respect to $\theta$, we obtain 
\[
\partial_\phi X^1=\chi\partial_\phi A(\phi,\theta).
\]
Setting the above equation in (2), it follows
\[
 r^2\sin^2\theta\partial_r X^2+ \frac{1}{\chi}\partial_\phi A(\phi,\theta)=0,
\]
Integrating the above equation with respect to $r$ implies
\begin{align}\label{rA02;}
X^2=\frac{\chi}{r} \csc^2\theta \partial_\phi A(\phi,\theta)+B(\phi,\theta).
\end{align}
From (\ref{rA02;}) and (5), we deduce 
\begin{align}\label{rA03;}
X^3=-\frac{\chi}{r} (-2\cot\theta A(\phi,\theta)+\partial_\theta A(\phi,\theta))-\sin^2\theta\int \partial_\theta B(\phi,\theta)d\phi+C(r,\theta).
\end{align}
Putting (\ref{rA01;}) and (\ref{rA03;}) in (3) yields
\begin{align}\label{rA05;}
-2\cot\theta A(\phi,\theta)+2\partial_\theta A(\phi,\theta)+\int \frac{1}{2\chi}\partial_\theta\rho(r,\theta)dr+r^2{\chi}\partial_rC(r,\theta)=0.
\end{align}
Differentiating (\ref{rA05;}) with respect to $\phi$, we obtain 
\begin{align*}
-\cot\theta\partial_\phi A(\phi,\theta)+\partial_\phi\partial_\theta A(\phi,\theta)=0.
\end{align*}
The above equation has the following solution: 
\begin{align}\label{rA07;}
A(\phi,\theta)=\sin\theta\int A_1(\phi)d\phi+A_2(\theta).
\end{align}
Setting (\ref{rA07;}) in (\ref{rA01;}), (\ref{rA02;}) and (\ref{rA03;}), we have
\begin{align}
X^1=&\chi\big[\int \frac{1}{2\chi}\rho(r,\theta)dr+\sin\theta\int A_1(\phi)d\phi+A_2(\theta)\big],\label{01}\\
X^2=&\frac{\chi}{r} \csc\theta  A_1(\phi)+B(\phi,\theta),\label{02}\\
X^3=&\frac{\chi}{r} (\cos\theta\int A_1(\phi)d\phi+2A_2(\theta)\cot\theta -\frac{dA_2(\theta)}{d\theta})-\sin^2\theta\int \partial_\theta B(\phi,\theta)d\phi+C(r,\theta).\label{03}
\end{align}
Applying (\ref{01}),  (\ref{03}) in (6), we get
\begin{align*}
&\chi\big[\int \frac{1}{2\chi}\rho(r,\theta)dr+A_2(\theta)+2\frac{dA_2(\theta)}{d\theta}\cot\theta -2A_2(\theta)\csc^2\theta -\frac{d^2A_2(\theta)}{d\theta^2}\big]\\
&+r(-\sin2\theta\int \partial_\theta B(\phi,\theta)d\phi-\sin^2\theta\int \partial^2_\theta B(\phi,\theta)d\phi+\partial_\theta C(r,\theta))=\frac{r}{2}\rho(r,\theta).
\end{align*}
Applying $\partial_\phi$ to both sides of the above equation, we have
\begin{align*}
r(-\sin2\theta \partial_\theta B(\phi,\theta)-\sin^2\theta \partial^2_\theta B(\phi,\theta))=0,
\end{align*}
which gives us 
\begin{align*}
 \partial^2_\theta B(\phi,\theta))+2\cot\theta\partial_\theta B(\phi,\theta)=0,
\end{align*}
that has the following solution:
\begin{align*}
B(\phi,\theta))=-B_1(\phi)\cot\theta+B_2(\phi).
\end{align*}
Putting the last equation  in (\ref{01}) and  (\ref{03}), we have
\begin{align}
X^1=&\chi\big[\int \frac{1}{2\chi}\rho(r,\theta)dr+\sin\theta\int A_1(\phi)d\phi+A_2(\theta)\big],\label{001}\\
X^3=&\frac{\chi}{r} (\cos\theta\int A_1(\phi)d\phi+2A_2(\theta)\cot\theta -\frac{dA_2(\theta)}{d\theta})-\int B_1(\phi)d\phi+C(r,\theta).\label{003}
\end{align}
Setting the above relations in (4), we get
\begin{align*}
&\chi\Big[\int \frac{\sin\theta}{\chi}\rho(r,\theta)dr+2\int A_1(\phi)d\phi+2\sin\theta A_2(\theta)+4A_2(\theta)\cos\theta\cot\theta -2\cos\theta\frac{dA_2(\theta)}{d\theta}\\
&+2\frac{dA_1(\phi)}{d\phi}\Big]-2r\cos\theta\int B_1(\phi)d\phi+2r\cos\theta C(r,\theta)-2r\cos\theta\frac{dB_1(\phi)}{d\phi}+2r\sin\theta\frac{dB_2(\phi)}{d\phi}=\rho(r,\theta)r\sin\theta.
\end{align*}
Differentiating the last equation with respect to $\phi$, we obtain 
\begin{align}\label{r801,;}
&\chi(\frac{d^2A_1(\phi)}{d\phi^2}+A_1(\phi))+r(\sin\theta\frac{d^2B_2(\phi)}{d\phi^2}-\cos\theta\frac{d^2B_1(\phi)}{d\phi^2}-\cos\theta B_1(\phi))=0.
\end{align}
Applying $\partial^2_r$ to both sides of the above equation, we deduce
\begin{align*}
	&\frac{d^2A_1(\phi)}{d\phi^2}+A_1(\phi)=0,
\end{align*}
which has the following solution:
\begin{align}\label{r80000,;}
A_1(\phi)=a_1\cos\phi+a_2\sin\phi, 
\end{align}
where $a_1$ and $ a_2$ are constants.
From (\ref{r801,;}) and (\ref{r800,;}), it follows
\begin{align*}
	&B_1(\phi)+\frac{d^2B_1(\phi)}{d\phi^2}=0,\ \ \ \ \ \frac{d^2B_2(\phi)}{d\phi^2}=0,
\end{align*}
which have the following solutions:
\begin{align}\label{r800,;}
B_1(\phi)=B_1\cos\phi+B_2\sin\phi, 
\end{align}
and 
\begin{align}\label{r8000,;}
B_2(\phi)=b_1\phi+b_2,
\end{align}
where $B_1, B_2, b_1$ and $ b_2$ are constants.
Setting (\ref{r80000,;}), (\ref{r800,;}) and (\ref{r8000,;})  in (\ref{001}), (\ref{02})  and  (\ref{003}), we have
\begin{align}
X^1=&\chi\big[\int \frac{1}{2\chi}\rho(r,\theta)dr+\sin\theta(a_1\sin\phi-a_2\cos\phi)+A_2(\theta)\big],\label{010}\\
X^2=&\frac{\chi}{r} \csc\theta  (a_1\cos\phi+a_2\sin\phi)-(B_1\cos\phi+B_2\sin\phi)\cot\theta+b_1\phi+b_2,\label{020}\\
X^3=&\frac{\chi}{r} (\cos\theta(a_1\sin\phi-a_2\cos\phi)+2A_2(\theta)\cot\theta -\frac{dA_2(\theta)}{d\theta})\label{030}\\
&-(B_1\sin\phi-B_2\cos\phi)+C(r,\theta).\nonumber
\end{align}
On the other hand, (4) and (6) imply
\begin{align}\label{N50}
\partial_\theta X^3-\cot\theta X^3=\partial_\phi X^2.
\end{align}
Setting (\ref{020}) and (\ref{030}) in the last equation gives
\begin{align}\label{21'}
&\frac{\chi}{r} (3\cos\theta \frac{dA_2(\theta)}{d\theta}-2\csc\theta A_2(\theta)-\sin\theta\frac{d^2A_2(\theta)}{d\theta^2}-2A_2(\theta)\cos\theta\cot\theta) +\sin\theta\partial_\theta C(r,\theta)\\
&+B3\cos\theta-C(r,\theta)\cos\theta=b_1\sin\theta.\nonumber
\end{align}
Differentiating the above equation with respect to $r$, we obtain 
\begin{align*}
&-\frac{1}{r^2\chi} (3\cos\theta \frac{dA_2(\theta)}{d\theta}-2\csc\theta A_2(\theta)-\sin\theta\frac{d^2A_2(\theta)}{d\theta^2}-2A_2(\theta)\cos\theta\cot\theta) \\
&+\sin\theta\partial_r\partial_\theta C(r,\theta)-\partial_rC(r,\theta)\cos\theta=0.
\end{align*}
Applying $-{r^2\chi} $ to the above equation, we get
\begin{align}\label{23}
&3\cos\theta \frac{dA_2(\theta)}{d\theta}-2\csc\theta A_2(\theta)-\sin\theta\frac{d^2A_2(\theta)}{d\theta^2}-2A_2(\theta)\cos\theta\cot\theta \\
&+{r^2\chi}(\partial_rC(r,\theta)\cos\theta-\sin\theta\partial_r\partial_\theta C(r,\theta))=0.\nonumber
\end{align}
Differentiating the above equation with respect to $r$, we have
\begin{align*}
&\frac{r(2-3kr^2)}{\chi}(\partial_rC(r,\theta)\cos\theta-\sin\theta\partial_r\partial_\theta C(r,\theta))+{r^2\chi}(\partial^2_rC(r,\theta)\cos\theta-\sin\theta\partial^2_r\partial_\theta C(r,\theta))=0,
\end{align*}
which gives
\begin{align*}
&\cos\theta\big({r(2-3kr^2)}\partial_rC(r,\theta)+r^2({\chi^2})\partial^2_rC(r,\theta)\big)\\
&-\sin\theta\big({r(2-3kr^2)}\partial_r\partial_\theta C(r,\theta)+r^2({\chi^2})\partial^2_r\partial_\theta C(r,\theta)\big)=0.
\end{align*}
Considering 
\begin{align}\label{AW1}
\bar C(r,\theta)={r(2-3kr^2)}\partial_rC(r,\theta)+r^2({\chi^2})\partial^2_rC(r,\theta),
\end{align}
in the above equation, we get
\begin{align*}
&\cos\theta\bar C(r,\theta)-\sin\theta\partial_\theta\bar C(r,\theta)=0,
\end{align*}
which has the following solution:
\begin{align*}
&\bar C(r,\theta)=C_1(r)\sin\theta.
\end{align*}
Setting the last equation in (\ref{AW1}), it follows that
\begin{align}\label{AW2}
{r(2-3kr^2)}\partial_rC(r,\theta)+r^2({\chi^2})\partial^2_rC(r,\theta)=C_1(r)\sin\theta.
\end{align}
Putting $\tilde C(r,\theta)=\partial_rC(r,\theta)$ in the above equation, we have 
\begin{align}\label{AW2'}
\partial_r\tilde C(r,\theta)+\frac{(2-3kr^2)}{r(1-kr^2)}\tilde C(r,\theta)=\frac{1}{r^2({\chi^2})}C_1(r)\sin\theta.
\end{align}
which has the following solution:
\begin{align}\label{024}
C(r,\theta)=\sin\theta\int\frac{1}{r^2\chi}\int \frac{C_1(r)}{\chi}drdr-C_2(\theta)\frac{\chi}{r}+C_3(\theta).
\end{align}
Setting (\ref{23}) in (\ref{21'}), we get
 \begin{align*}
&{r({\chi^2})}(\sin\theta\partial_r\partial_\theta C(r,\theta)-\cos\theta\partial_rC(r,\theta))+\sin\theta\partial_\theta C(r,\theta)\\
&-C(r,\theta)\cos\theta-b_1\sin\theta=0.
\end{align*}
The above two equations imply
\[
\frac{dC_3(\theta)}{d\theta}-\cot\theta C_3(\theta)=b_1,
\]
which has the following solution:
\begin{align}\label{0301}
C_3(\theta)=b_1\sin\theta\ln(\csc\theta-\cot\theta)+C_4\sin\theta.
\end{align}
Putting (\ref{024}) and (\ref{0301}) in (\ref{030}), we have 
\begin{align}\label{0300}
&X^3=\frac{\chi}{r} (\cos\theta(a_1\sin\phi-a_2\cos\phi)+2A_2(\theta)\cot\theta -\frac{dA_2(\theta)}{d\theta})\\
&-(B_1\sin\phi-B_2\cos\phi)+\sin\theta\int\frac{1}{r^2\chi}\int \frac{C_1(r)}{\chi}drdr-C_2(\theta)\frac{\chi}{r}\nonumber\\
&+b_1\sin\theta\ln(\csc\theta-\cot\theta)+C_4\sin\theta.\nonumber
\end{align}
Substituting (\ref{010}) and (\ref{0300}) in (3) implies 
\begin{align}\label{03000}
&-2A_2(\theta)\cot\theta+2\frac{dA_2(\theta)}{d\theta}+\sin\theta\int \frac{C_1(r)}{\chi}dr+C_2(\theta)\\
&+\int \frac{1}{2\chi}\partial_\theta \rho(r,\theta)dr=0.\nonumber
\end{align}
Differentiating the above relation with respect to $r$ gives
\[
\partial_\theta \rho(r,\theta)=-2C_1(r)\sin\theta.
\]
Hence we get 
\begin{align}\label{p1}
\rho(r,\theta)=2C_1(r)\cos\theta+\rho_1(r).
\end{align}
Plugging (\ref{p1}) in (\ref{03000}), it follows
\begin{align}\label{M1}
&-2A_2(\theta)\cot\theta+2\frac{dA_2(\theta)}{d\theta}+C_2(\theta)=0.
\end{align}
The above differential equation has the following solution:
\begin{align}\label{M11}
&A_2(\theta)=\sin\theta(-\frac{1}{2}\int \csc\theta C_2(\theta){d\theta}+C_5).
\end{align}
Putting (\ref{M11}) in (\ref{010}), (\ref{020})  and  (\ref{0300}), we have 
\begin{align}
X^1=&\chi\big[\int \frac{1}{\chi}(C_1(r)\cos\theta+\frac{1}{2}\rho_1(r))dr+\sin\theta(a_1\sin\phi-a_2\cos\phi)\label{010'}\\
&+\sin\theta(-\frac{1}{2}\int \csc\theta C_2(\theta){d\theta}+C_5)\big],\nonumber\\
X^2=&\frac{\chi}{r} \csc\theta  (a_1\cos\phi+a_2\sin\phi)-(B_1\cos\phi+B_2\sin\phi)\cot\theta+b_1\phi+b_2,\label{N02000}\\
X^3=&\frac{\chi}{r} (\cos\theta(a_1\sin\phi-a_2\cos\phi+C_5)-\frac{1}{2}\cos\theta\int \csc\theta C_2(\theta){d\theta}\label{0400}\\
&-\frac{1}{2}C_2(\theta))-(B_1\sin\phi-B_2\cos\phi)+\sin\theta\int\frac{1}{r^2\chi}\int \frac{C_1(r)}{\chi}drdr\nonumber\\
&+b_1\sin\theta\ln(\csc\theta-\cot\theta)+C_4\sin\theta.\nonumber
\end{align}
Setting (\ref{p1}), (\ref{010'}) and (\ref{0400}) in (6), we get 
\begin{align*}
&\frac{\chi}{r}\big[\int \frac{1}{\chi}(C_1(r)\cos\theta+\frac{1}{2}\rho_1(r))dr-\frac{1}{2}\cot\theta C_2(\theta)-\frac{1}{2}\frac{dC_2(\theta)}{d\theta}]\\
&+\cos\theta[\int\frac{1}{r^2\chi}\int \frac{C_1(r)}{\chi}drdr+b_1\ln(\csc\theta-\cot\theta)+C_4-C_1(r)]+b_1-\frac{1}{2}\rho_1(r)=0.\nonumber
\end{align*}
Differentiating the last relation with respect to $r$, we have
\begin{align}\label{N37}
&-\frac{1}{2r^2\chi}\big[\int \frac{1}{\chi}\rho_1(r)dr-\cot\theta C_2(\theta)-\frac{dC_2(\theta)}{d\theta}]\\
&+\frac{1}{r}(\cos\theta C_1(r)+\frac{1}{2}\rho_1(r))-\cos\theta\frac{dC_1(r)}{dr}-\frac{1}{2}\frac{\partial\rho_1(r)}{\partial r}=0.\nonumber
\end{align}
Applying $\frac{\partial}{\partial{\theta}}$ to  the above equation implies
\begin{align}\label{N38}
&-\frac{1}{2r^2\chi}\big[\csc^2\theta C_2(\theta)-\cot\theta\frac{dC_2(\theta)}{d\theta}-\frac{d^2C_2(\theta)}{d\theta^2}]-\frac{1}{r}\sin\theta C_1(r)+\sin\theta\frac{dC_1(r)}{dr}=0,
\end{align}
which gives us 
\begin{align*}
&\csc^3\theta C_2(\theta)-\cot\theta\csc\theta\frac{dC_2(\theta)}{d\theta}-\csc\theta\frac{d^2C_2(\theta)}{d\theta^2}={2r^2\chi}(-\frac{1}{r} C_1(r)+\frac{dC_1(r)}{dr}).\nonumber
\end{align*}
The above equation holds if and only if
\begin{align*}
&\csc^3\theta C_2(\theta)-\cot\theta\csc\theta\frac{dC_2(\theta)}{d\theta}-\csc\theta\frac{d^2C_2(\theta)}{d\theta^2}=N={2r^2\chi}(-\frac{1}{r} C_1(r)+\frac{dC_1(r)}{dr}),\nonumber
\end{align*}
where $N$ is constant. Now we focus on the differential equation
\begin{align}\label{FD}
{2r^2\chi}(-\frac{1}{r} C_1(r)+\frac{dC_1(r)}{dr})=N.
\end{align}
As the above equation  is $\kappa$-dependent,  we consider the two cases:

{\bf Case 1. $\kappa\neq 0$}\\
In this case, (\ref{FD}) has the following solution:
\begin{align}\label{N39}
 C_1(r)=r[-\frac{N}{4}\big(\frac{\chi}{r^2}+k \tanh^{-1}(\frac{1}{\chi})\big)+\mathcal{A}],
\end{align}
where $\mathcal{A}$ is a constant. Also,  equation 
\begin{align*}
&\csc^3\theta C_2(\theta)-\cot\theta\csc\theta\frac{dC_2(\theta)}{d\theta}-\csc\theta\frac{d^2C_2(\theta)}{d\theta^2}=N,
\end{align*}
has the solution 
\begin{align}\label{N41}
C_2(\theta)=\frac{N}{2}\sin\theta-S\cot\theta+M\csc\theta,
\end{align}
where  $M$ and  $S$ are constants. 
Using (\ref{N39}) and (\ref{N41}) in (\ref{0400}), we have
\begin{align}
X^3=&\frac{\chi}{r} [\cos\theta(a_1\sin\phi-a_2\cos\phi+C_5)-\frac{1}{4}N(\theta\cos\theta+\sin\theta)-\frac{1}{2}M\sin\theta]\label{N0400}\\
&+\sin\theta\int\frac{1}{r^2\chi}\int \frac{r}{\chi}[-\frac{N}{4}\big(\frac{\chi}{r^2}+k \tanh^{-1}(\frac{1}{\chi})\big)+\mathcal{A}]drdr\nonumber\\
&-(B_1\sin\phi-B_2\cos\phi)+b_1\sin\theta\ln(\csc\theta-\cot\theta)+C_4\sin\theta.\nonumber
\end{align}
Setting (\ref{N02000}) and (\ref{N0400}) in (\ref{N50}), we obtain
\[
\frac{N}{4}(-\cos\theta+\theta\csc\theta)-C_5\csc\theta=0,
\]
which gives
\[
\frac{N}{4}(-\cos\theta\sin\theta+\theta)=C_5,
\]
thus $N=C_5=0$. Hence (\ref{N39}) and (\ref{N41}) reduce to
\begin{align}\label{N39'}
C_1(r)=r\mathcal{A},
\end{align}
and
\begin{align}\label{N41'}
C_2(\theta)=-S\cot\theta+M\csc\theta.
\end{align}
Replacing (\ref{N39'}) in (\ref{p1}), we have
\begin{align}\label{p12}
\rho(r,\theta)=2r\mathcal{A}\cos\theta+\rho_1(r).
\end{align}
Using (\ref{N39'}) and (\ref{N41'}) in (\ref{010'}), (\ref{N02000}) and (\ref{N0400}), we get
\begin{align}
X^1=&-\frac{\chi^2}{k}\mathcal{A}\cos\theta+\chi\big[\int \frac{1}{2\chi}\rho_1(r)dr+\sin\theta(a_1\sin\phi-a_2\cos\phi)\label{N010'}\\
&+\frac{1}{2}M\cos\theta-\frac{1}{2}S\big],\nonumber\\
X^2=&\frac{\chi}{r} \csc\theta  (a_1\cos\phi+a_2\sin\phi)-(B_1\cos\phi+B_2\sin\phi)\cot\theta+b_1\phi+b_2,\label{N020'}\\
X^3=&\frac{\chi}{r} [\cos\theta(a_1\sin\phi-a_2\cos\phi)-\frac{1}{2}M\sin\theta]+\frac{1}{rk}\mathcal{A}\sin\theta-(B_1\sin\phi-B_2\cos\phi)\label{N0400'}\\
&+b_1\sin\theta\ln(\csc\theta-\cot\theta)+C_4\sin\theta.\nonumber
\end{align}
Putting (\ref{N010'}) and (\ref{N0400'}) in (6) implies
\begin{align}\label{N51}
\frac{\chi}{r}\big[\int \frac{1}{2\chi}\rho_1(r)dr-\frac{1}{2}S\big]+b_1\cos\theta\ln(\csc\theta-\cot\theta)+b_1+C_4\cos\theta=\frac{1}{2}\rho_1(r).
\end{align}
Differentiating  the last equation  with respect to $\theta$, it follows that
\[
-b_1\sin\theta\ln(\csc\theta-\cot\theta)+b_1\cot\theta-C_4\sin\theta=0.
\]
Considering $\theta=\frac{\pi}{2}$ in the above equation, we deduce 
\[
C_4=b_1=0.
\]
Hence (\ref{N51}) implies
\begin{align}\label{N52}
\int \frac{1}{2\chi}\rho_1(r)dr-\frac{1}{2}S-\frac{r}{2\chi}\rho_1(r)=0.
\end{align}
Differentiating  the above equation  with respect to $r$ gives
\begin{align}
\frac{d\rho_1(r)}{dr}+\frac{kr}{{\chi^2}}\rho_1(r)=0,
\end{align}
that has the following solution:
\begin{align}\label{N53}
\rho_1(r)=\mathcal{B}\chi,
\end{align}
where $\mathcal{B}$ is constant. From (\ref{N52}) and (\ref{N53}), we get 
\[
S=0.
\]
Therefore, from (\ref{N010'})-(\ref{N0400'}) we conclude
\begin{align}
X^1=&-\frac{\chi^2}{k}\mathcal{A}\cos\theta+\chi\big[ \frac{1}{2}\mathcal{B}r+\sin\theta(a_1\sin\phi-a_2\cos\phi)+\frac{1}{2}M\cos\theta\big]\label{N010''},\\
X^2=&\frac{\chi}{r} \csc\theta  (a_1\cos\phi+a_2\sin\phi)-(B_1\cos\phi+B_2\sin\phi)\cot\theta+b_2,\label{N020''}\\
X^3=&\frac{\chi}{r} [\cos\theta(a_1\sin\phi-a_2\cos\phi)-\frac{1}{2}M\sin\theta]+\frac{1}{rk}\mathcal{A}\sin\theta-(B_1\sin\phi-B_2\cos\phi)\label{N0400''}.
\end{align}

{\bf Case 2. $\kappa=0$}\\
In this case, (\ref{FD}) has the following solution:
\begin{align}\label{FD1}
C_1(r)=-\frac{N}{4r}+\mathcal{A}r.
\end{align}
Using (\ref{FD1}) and (\ref{N41}) in (\ref{0400}), we have
\begin{align}
X^3=&\frac{1}{r} [\cos\theta(a_1\sin\phi-a_2\cos\phi+C_5)-\frac{1}{4}N(\theta\cos\theta+\sin\theta)-\frac{1}{2}M\sin\theta]+\frac{N}{4r}(\ln r+1)\sin\theta\label{FN0400}\\
&+\frac{\mathcal{A}}{2}r\sin\theta-(B_1\sin\phi-B_2\cos\phi)+b_1\sin\theta\ln(\csc\theta-\cot\theta)+C_4\sin\theta.\nonumber
\end{align}
Setting (\ref{N02000}) (when $\kappa=0$) and (\ref{FN0400}) in (\ref{N50}), similarly to the previous case we can conclude $N=C_5=0$. So, we have the relations (\ref{N39'}), (\ref{N41'}) and (\ref{p12}). Using these equations, $X^1$ and $X^3$  reduce to the following
\begin{align}
X^1=&\frac{\mathcal{A}r^2}{2}\cos\theta+\frac{1}{2}\int\rho_1(r)dr+\sin\theta(a_1\sin\phi-a_2\cos\phi)+\frac{1}{2}M\cos\theta-\frac{1}{2}S,\label{FN010'}\\
X^3=&\frac{1}{r} [\cos\theta(a_1\sin\phi-a_2\cos\phi)-\frac{1}{2}M\sin\theta]-(B_1\sin\phi-B_2\cos\phi)+\frac{\mathcal{A}}{2}r\sin\theta\label{FFN0400}\\
&+b_1\sin\theta\ln(\csc\theta-\cot\theta)+C_4\sin\theta.\nonumber
\end{align}
According the above two equations along with (6), we can write 
\begin{align}\label{FN51}
 \frac{1}{2}\int\rho_1(r)dr-\frac{1}{2}S+b_1r\cos\theta\ln(\csc\theta-\cot\theta)+b_1r+C_4r\cos\theta=\frac{1}{2}\rho_1(r)r.
\end{align}
Applying $\frac{d}{d{\theta}}$ to  (\ref{FN51}) and setting  $\theta=\frac{\pi}{2}$  implies $C_4=b_1=0$. Hence, the above equation yields 
\begin{align}\label{FFN51}
 \frac{1}{2}\int\rho_1(r)dr-\frac{1}{2}S=\frac{1}{2}\rho_1(r)r.
\end{align}
Differentiating the last relation with respect to $r$, we deduce $\frac{d\rho_1(r)}{d{r}}=0$ and consequently, $\rho_1(r)=\mathcal{B}$, where $\mathcal{B}$ is constant. This and (\ref{FFN51}) give $S=0$. Therefore, (\ref{N02000}),  (\ref{FN010'}) and (\ref{FFN0400}) reduce to
\begin{align}
X^1=&\frac{\mathcal{A}r^2}{2}\cos\theta+\frac{1}{2}\mathcal{B}r+\sin\theta(a_1\sin\phi-a_2\cos\phi)+\frac{1}{2}M\cos\theta,\label{FFN010'}\\
X^2=&\frac{1}{r} \csc\theta  (a_1\cos\phi+a_2\sin\phi)-(B_1\cos\phi+B_2\sin\phi)\cot\theta+b_2,\label{FFN020''}\\
X^3=&\frac{1}{r} [\cos\theta(a_1\sin\phi-a_2\cos\phi)-\frac{1}{2}M\sin\theta]-(B_1\sin\phi-B_2\cos\phi)+\frac{\mathcal{A}}{2}r\sin\theta.\label{FFFN0400}\end{align}
\begin{theorem}\label{A102}
The vector fields $X=X^1\partial_r+X^2\partial_\phi+X^3\partial_\theta$ are conformal vector fields on FLRW with a conformal factor $\rho( r, \theta)$ if and only if $X^i, i=1, 2, 3$ satisfy (\ref{N010''})-(\ref{N0400''}) (for $\kappa\neq 0$) and  (\ref{FFN010'})-(\ref{FFFN0400}) (for $\kappa=0$), where $\rho$ is given by $\rho(r,\theta)=2r\mathcal{A}\cos\theta+\mathcal{B}\chi$ (for $\kappa\neq 0$) and $\rho(r,\theta)=2r\mathcal{A}\cos\theta+\mathcal{B}$ (for $\kappa=0$) and $\mathcal{A}$, $\mathcal{B}$ are constant.
\end{theorem}
Now, we write the explicit solutions for CKVF's of the FLRW space-times.
Setting 
$C_2=1$ in (\ref{AAAA001;})-(\ref{AAAA003;}) and (\ref{pphi377})-(\ref{ppphi377}),  $B_2=1$ in (\ref{qL00i})-(\ref{qL00iii}) and 
$B_1=-1$ in (\ref{N010''})-(\ref{N0400''}) and (\ref{FFN010'})-(\ref{FFFN0400}), 
 the first CKVF is obtained as follows:
\[
\xi^1=\sin\phi\partial_{\theta}+\frac{\cos\phi}{\tan\theta}\partial_\phi.
\]
Putting $C_1=1$ in  (\ref{AAAA001;})-(\ref{AAAA003;}) and (\ref{pphi377})-(\ref{ppphi377}), $B_1=$ in (\ref{qL00i})-(\ref{qL00iii}) and $B_2=1$ in (\ref{N010''})-(\ref{N0400''}) and (\ref{FFN010'})-(\ref{FFFN0400}), we find
the second CKVF as 
\[
\xi^2=\cos\phi\partial_\theta-\frac{\sin\phi}{\tan\theta}\partial_\phi.
\]
In the same way, by setting specific values instead of constant coefficients in (\ref{AAAA001;})-(\ref{AAAA003;}), (\ref{pphi377})-(\ref{ppphi377}),  (\ref{qL00i})-(\ref{qL00iii}), (\ref{N010''})-(\ref{N0400''}) and (\ref{FFN010'})-(\ref{FFFN0400}), we find
\begin{align}
\xi^3=&\partial_\phi,\nonumber\\
\xi^4=&\chi\{\sin\theta\cos\phi\partial_r-\frac{1}{r}\frac{\sin\phi}{\sin\theta}\partial_\phi+\frac{1}{r}\cos\theta\cos\phi\partial_\theta\},\nonumber\\
\xi^5=&\chi\{\sin\theta\sin\phi\partial_r+\frac{1}{r}\frac{\cos\phi}{\sin\theta}\partial_\phi+\frac{1}{r}\cos\theta\sin\phi\partial_\theta\},\nonumber\\
\xi^6=&\chi\{-\cos\theta\partial_r+\frac{1}{r}{\sin\theta}\partial_\theta\},\nonumber\\
\xi^7=&\frac{r}{2}\chi\partial_r,\nonumber\\
\xi^8=&-\frac{\chi^2}{\kappa}\cos\theta\partial_r+\frac{1}{r\kappa}\sin\theta\partial_\theta,\nonumber\\
\xi^9=&\frac{r}{2}\{r\cos\theta\partial_r+\sin\theta\partial_\theta\}.\nonumber
\end{align}
It is easy to see that $\pounds_{\xi^i}g=0$, for $ i=1,\cdots, 6$, and
\begin{align*}
\pounds_{\xi^7}g_{\nu\lambda}
=
\begin{pmatrix}
\frac{a^2}{{\chi}}&0&0&0\\
0&{r^2a^2{\chi}\sin^2\theta}&0&0\\
0&0&{r^2a^2{\chi}}&0\\
0&0&0&	0
\end{pmatrix},\ \ \ \ \ \pounds_{\xi^8}g_{\nu\lambda}
=
\begin{pmatrix}
\frac{2ra^2\cos^2\theta}{\chi}&0&0&0\\
0&2r^3a^2\cos\theta\sin^2\theta&0&0\\
0&0&2r^3a^2\cos\theta&	0\\
0&0&0&0
\end{pmatrix},\ \ \ \ \ 
\end{align*} 
\begin{align*}
\pounds_{\xi^9}g_{\nu\lambda}
=	\begin{pmatrix}
{2ra^2\cos^2\theta}&0&0&0\\
0&2r^3a^2\cos\theta\sin^2\theta&0&0\\
0&0&2r^3a^2\cos\theta&	0\\
0&0&0&0
\end{pmatrix}, 
\end{align*}
where $a\neq 0$ is constant, thus  $\xi^i, i=1,\cdots,6$
are  Killing vector fields and  $\xi^i, i=7,8,9$ are non-Killing vector fields such that $\rho(r,\theta)=\chi$ is the conformal factor of $\xi^7$, and $\rho(r,\theta)=2r\cos\theta$ is the conformal factor of $\xi^8$ and $\xi^9$.
\begin{corollary}\label{AM6}
According to the explanation above,
there are indeed $9$ CKVF's on FLRW space-times such that, $\xi^7$ is the proper  homothetic vector field for $\kappa=0$ and the proper conformal vector field for $\kappa\neq 0$. Moreover, $\xi^8$ and $\xi^9$ are proper conformal vector fields.  
\end{corollary}
\begin{remark}
The number of CKVF's obtained in Corollary \ref{AM6}, confirms the dimension of the Lie algebras of conformal, homothetic, and Killing vector fields found in \cite{hall}. 
\end{remark}
\section{Statistical structures on FLRW space-time}
\begin{definition}
	\label{def1} 
	Let $\nabla$ be an affine connection and $g$ be a pseudo-Riemannian metric. The pair $(\nabla ,g)$ is called  Codazzi-coupled if 
	the non-metricity $Q = {\nabla}g$ is totally symmetric; namely the Codazzi equations hold:
	\begin{equation}
	{\nabla}_\lambda g_{\mu\nu} = {\nabla}_\mu g_{\lambda\nu}, \qquad (= {\nabla}_\nu g_{\lambda\mu}),  \label{S200}
	\end{equation}%
\end{definition}
In local coordinates, the non-metricity $Q$ has the following form
\begin{align}\label{8}
Q_{\lambda\mu\nu}=Q_{\mu\lambda\nu}=Q_{\nu\lambda\mu}.
\end{align}
Using  the affine connection $\nabla$ and the Levi-Civita connection $\hat{\nabla}$, we define the different tensor $K$ as
   $K=\nabla-\hat{\nabla}$, which in local coordinates has the expression
\begin{align}
&K^\lambda_{\mu\nu}=\Gamma^\lambda_{\mu\nu}-\hat\Gamma^\lambda_{\mu\nu}.
\end{align}
Applying the above equation in (\ref{3999}) and (\ref{3}), it follows
\begin{align*}
&T^\lambda_{\mu\nu}=K^\lambda_{\mu\nu}-K^\lambda_{\nu\mu},
\end{align*}
and
\[
Q_{\lambda\mu\nu} =-K^{\sigma}_{\lambda\mu}g_{\nu\sigma}-K^{\sigma}_{\lambda\nu}g_{\mu\sigma}.
\]
Moreover, the Codazzi-coupled  pair $(\nabla,g)$ is called a statistical structure if $\nabla$ is torsion-free, i.e., $T=0$.
In particular,  it is a well known fact that if the non-metricity $Q$ is zero,  the torsion-free  connection $\nabla$ reduces to the Levi-Civita connection $\hat{\nabla}$.
 For the statistical structure $(\nabla, g)$, it is obvious that
\begin{align*}
K^\lambda_{\mu\nu}=K^\lambda_{\nu\mu},\ \ \ \ \ \ K^\lambda_{\mu\nu}g_{\lambda\kappa}=K^\lambda_{\mu\kappa}g_{\lambda\nu}, \ \ \ \ Q_{\lambda\mu\nu} =-2K^{\sigma}_{\lambda\mu}g_{\nu\sigma}.
\end{align*}

In the sequel, we focus on creating a statistical structure on FLRW space-time and investigate 
 the conformal symmetry of $K$, i.e, we study the condition $\pounds_{\xi^i}K^\mu_{\nu\lambda}=0$, for $i=1, \cdots, 9$, where $\pounds_{\xi^i}$ is the Lie derivative with respect to the vector fields $\xi^i$ given by Corollary \ref{AM6}. Indeed, this condition is equivalent to obtaining the statistical connections satisfying $\pounds_{\xi^i}\Gamma^\mu_{\nu\lambda}=\pounds_{\xi^i}\hat{\Gamma}^\mu_{\nu\lambda}$, where $\hat{\Gamma}^\mu_{\nu\lambda}$ is the Levi-Civita connection of FLRW. So, first we study the Lie derivative of the Levi-Civita connection with respect to $\xi^i$.

Using the explicit coordinate description of FLRW metric given by (\ref{1}) and the
well-known Koszul formula, one can calculate the non-zero components of the Levi-Civita
connection in the local coordinates $(r,\phi,\theta,t)$ as follows:
\begin{align}
&\hat\Gamma^t_{tt}=\frac{\dot N(t)}{N(t)},\ \ \ \ \hat\Gamma^t_{\theta\theta}=\frac{\dot a(t)a(t)r^2}{N^2(t)},\ \ \ \ \hat\Gamma^t_{rr}=\frac{\dot a(t)a(t)}{(\chi^2)N^2(t)},\ \ \ \ 
\label{K9}\\
&\hat\Gamma^t_{\phi \phi}=\frac{\dot a(t)a(t)r^2\sin^2\theta}{N^2(t)}, \ \ \ \
\hat\Gamma^r_{tr}=\hat\Gamma^r_{rt}=\hat\Gamma^\theta_{\theta t}=\hat\Gamma^\theta_{t\theta}=\hat\Gamma^\phi_{ t\phi}=\hat\Gamma^\phi_{\phi t}=\frac{\dot a(t)}{a(t)},\nonumber\\
&
\hat\Gamma^r_{rr}=\frac{\kappa r}{\chi^2},\ \ \ \ \hat\Gamma^r_{\theta\theta}=-r(\chi^2),\ \ \ \ \hat\Gamma^r_{\phi \phi}\!=\!-r\sin^2\theta(\chi^2), \nonumber\\
& \hat\Gamma^\theta_{r\theta}\!=\!\hat\Gamma^\theta_{\theta r}\!=\!\hat\Gamma^\phi_{r\phi}\!=\!\hat\Gamma^\phi_{\phi r}\!=\!\frac{1}{r},\ \ \ \ \ \ \ \ \ \ \  \hat\Gamma^\theta_{\phi\phi}\!=-\sin\theta\cos\theta, \ \ \ \ \ \ \  \hat\Gamma^\phi_{\theta\phi}=\hat\Gamma^\phi_{\phi\theta}=\cot\theta.\nonumber
\end{align}
It is known that the infinitesimal transformation of the affine geometry is
defined by 
\begin{align}\label{AM2}
\pounds_X\Gamma^\mu_{\kappa\lambda}=X^\sigma\frac{\partial\Gamma^\mu_{\kappa\lambda}}{\partial x^\sigma}+\Gamma^\mu_{\sigma\lambda}\frac{\partial X^\sigma}{\partial x^\kappa}+\Gamma^\mu_{\kappa\sigma}\frac{\partial X^\sigma}{\partial x^\lambda}-\Gamma^\sigma_{\kappa\lambda}\frac{\partial X^\mu}{\partial x^\sigma}+\frac{\partial^2X^\mu}{\partial x^\kappa \partial x^\lambda}.
\end{align}
Using the above equation we obtain $\pounds_{\xi^i}\hat{\Gamma}^\mu_{\nu\lambda}=0$, $i=1, \cdots, 6$. So, $\pounds_{\xi^i}K^\mu_{\nu\lambda}=0$  if and only if $\pounds_{\xi^i}\Gamma^\mu_{\nu\lambda}=0$, for $i=1, \cdots, 6$, where $\Gamma^\mu_{\nu\lambda}$ are the components of a statistical connection. Therefore, we must obtain statistical connections with the property $\pounds_{\xi^i}\Gamma^\mu_{\nu\lambda}=0$, for $i=1, \cdots, 6$. Since $\xi^i$, $i=1, \cdots, 6$, are the generators of Killing vector fields, indeed, we study the Killing symmetry of statistical connections (It is known that an affine geometry is symmetric under a group action if the Lie derivative of the connection vanishes for all generating vector fields, i.e.,  $\pounds_X\Gamma^\mu_{\kappa\lambda}=0$). First, we do not impose any geometric postulates on the connection, that is, we start with a generic connection which has 64 components, all of which are functions of $(r,\phi,\theta, t)$, and we do not demand any flatness, metric-compatibility, or torsion-freeness. In the second step, we consider the statistical condition on them.

Setting $\xi^i, i=1,\cdots,6$ given by Corollary \ref{AM6} in (\ref{AM2}), it follows that $\pounds_{\xi^i}\Gamma^\mu_{\kappa\lambda}=0$ if and only if (see \cite{MH})
\begin{align}
&\Gamma^t_{tt}={\mathcal{K}_1(t)},\ \ \ \Gamma^t_{\theta\theta}={\mathcal{K}_2(t)}r^2,\ \  \Gamma^t_{rr}=\frac{{\mathcal{K}_2(t)}}{\chi^2},\ \ 
\Gamma^t_{\phi \phi}={\mathcal{K}_2(t)}r^2\sin^2\theta, \ \ 
\Gamma^r_{tr}=\Gamma^\theta_{t\theta}=\Gamma^\phi_{ t\phi}={\mathcal{K}_3(t)},\label{9}\\
&\Gamma^r_{rt}=\Gamma^\theta_{\theta t}=\Gamma^\phi_{\phi t}={\mathcal{K}_4(t)},\ \ \ \ \ \ \ \  \Gamma^r_{\phi\theta}=-\Gamma^r_{\theta\phi}={\mathcal{K}_5(t)}r^2\chi\sin\theta,\ \ \ \ \  \ \  \Gamma^\theta_{r\phi}=-\Gamma^\theta_{\phi r}=\frac{{\mathcal{K}_5(t)}\sin\theta}{\chi},\nonumber\\
&
\Gamma^\phi_{r\theta}=-\Gamma^\phi_{\theta r}=-\frac{{\mathcal{K}_5(t)}}{\chi\sin\theta},\ \ \ \ \ \ \ \ \ \  \Gamma^r_{rr}=\frac{\kappa r}{\chi^2},\ \ \ \ \ \ \ \ \Gamma^r_{\theta\theta}=-r(\chi^2),\ \   \ \ \  \Gamma^r_{\phi \phi}\!=\!-r\sin^2\theta(\chi^2), \nonumber\\
& \Gamma^\theta_{r\theta}\!=\!\Gamma^\theta_{\theta r}\!=\!\Gamma^\phi_{r\phi}\!=\!\Gamma^\phi_{\phi r}\!=\!\frac{1}{r},\ \ \ \ \ \ \ \   \Gamma^\theta_{\phi\phi}\!=-\sin\theta\cos\theta, \ \ \ \ \ \ \ \ \ \ \ \ \ \  \Gamma^\phi_{\theta\phi}=\Gamma^\phi_{\phi\theta}=\cot\theta.\nonumber
\ \ \ 
\end{align}
Now, we study the Codazzi condition on the connections given by (\ref{9}). According to (\ref{3}) 
we get
\begin{align*}
Q_{rrt} =& \frac{\mathcal{K}_2(t)N^2(t)-\mathcal{K}_4(t)a^2(t)}{\chi^2}=	Q_{rtr},\\
Q_{\phi\phi t} =& r^2\sin^2\theta(\mathcal{K}_2(t)N^2(t)-\mathcal{K}_4(t)a^2(t))=	Q_{\phi t\phi},\\
Q_{\theta\theta t} =& r^2(\mathcal{K}_2(t)N^2(t)-\mathcal{K}_4(t)a^2(t))=	Q_{\theta t\theta},\\
Q_{trr} =&\frac{2a(t)(\dot a(t)-\mathcal{K}_3(t)a(t))}{\chi^2},\\
Q_{t\phi\phi} =&{2a(t)r^2\sin^2\theta(\dot a(t)-\mathcal{K}_3(t)a(t))},\\
Q_{t\theta\theta} =&{2a(t)r^2(\dot a(t)-\mathcal{K}_3(t)a(t))},\\
Q_{ttt} =&{2N(t)(-\dot N(t)+\mathcal{K}_1(t)N(t))}.
\end{align*}
The above equations imply
\begin{align*}
&Q_{trr}-Q_{rtr}=\frac{2\dot a(t)a(t)-2\mathcal{K}_3(t)a^2(t)+\mathcal{K}_4(t)a^2(t)-\mathcal{K}_2(t)N^2(t)}{\chi^2},\\
& Q_{t\phi\phi}-Q_{\phi t\phi}=r^2\sin^2\theta(2\dot a(t)a(t)-2\mathcal{K}_3(t)a^2(t)+\mathcal{K}_4(t)a^2(t)-\mathcal{K}_2(t)N^2(t)),\\
&Q_{t\theta\theta}-Q_{\theta t\theta}=r^2(2\dot a(t)a(t)-2\mathcal{K}_3(t)a^2(t)+\mathcal{K}_4(t)a^2(t)-\mathcal{K}_2(t)N^2(t)).
\end{align*}
So, from the above equations, it folows that an FLRW space-time $g$ satisfies the Codazzi equations with the connections given by (\ref{9}) if and only if 
\begin{align}\label{10000}
2\dot a(t)a(t)-2\mathcal{K}_3(t)a^2(t)+\mathcal{K}_4(t)a^2(t)-\mathcal{K}_2(t)N^2(t)=0,
\end{align}
which demands
 the scale factor $a(t)$ to be given by
	\begin{align}\label{100}
	a(t)=\frac{ \sqrt{ e^{\int (-2\mathcal{K}_3(t)+ \mathcal{K}_4(t))dt}(\int\frac{e^{\int K_4(t)dt}\mathcal{K}_2(t)N^2(t)dt}{(e^{\int \mathcal{K}_3(t)}dt)^2}+c)}}{e^{\int(-2 \mathcal{K}_3(t)+\mathcal{K}_4(t)}dt}.
	\end{align}
To study the statistical connections, we consider the torsion-free condition on (\ref{9}). It implies 
\begin{align}\label{LLLLLL}
\mathcal{K}_3(t)=\mathcal{K}_4(t), \ \ \ \mathcal{K}_5(t)=0,
\end{align}
Considering the above relations, equations (\ref{9}) and (\ref{100}) reduce to the following 
\begin{align}
&\Gamma^t_{tt}={\mathcal{K}_1(t)},\ \ \ \ \Gamma^t_{\theta\theta}={\mathcal{K}_2(t)}r^2,\ \ \ \ \Gamma^t_{rr}=\frac{{\mathcal{K}_2(t)}}{\chi^2},\ \ \ \ 
\Gamma^t_{\phi \phi}={\mathcal{K}_2(t)}r^2\sin^2\theta, \ \ \ \
\label{9'}\\
&\Gamma^r_{tr}=\Gamma^\theta_{t\theta}=\Gamma^\phi_{ t\phi}=\Gamma^r_{rt}=\Gamma^\theta_{\theta t}=\Gamma^\phi_{\phi t}={\mathcal{K}_3(t)},\ \ \ \ \ \   \Gamma^r_{rr}=\frac{\kappa r}{\chi^2},\ \ \ \  \Gamma^r_{\phi \phi}\!=\!-r\sin^2\theta(\chi^2),\ \ \nonumber\\
& \Gamma^\theta_{r\theta}\!=\!\Gamma^\theta_{\theta r}\!=\!\Gamma^\phi_{r\phi}\!=\!\Gamma^\phi_{\phi r}\!=\!\frac{1}{r}, \ \ \ \ \Gamma^r_{\theta\theta}=-r(\chi^2),\ \  \Gamma^\theta_{\phi\phi}\!=-\sin\theta\cos\theta, \ \ \   \Gamma^\phi_{\theta\phi}=\Gamma^\phi_{\phi\theta}=\cot\theta, \nonumber
\end{align}
	\begin{align}\label{1000}
a(t)=\frac{ \sqrt{ e^{-\int K_3(t)dt}(\int\frac{\mathcal{K}_2(t)N^2(t)dt}{e^{\int \mathcal{K}_3(t)}dt}+c)}}{e^{-\int \mathcal{K}_3(t)}dt},
\end{align}
respectively.
Consequently we have:
\begin{theorem}\label{110}
 The connection given by (\ref{9'})
  together with the FLRW metric $g$ given by
  \begin{align*}
  g=-N^2(t)dt^2+\frac{ { e^{-\int K_3(t)dt}(\int\frac{\mathcal{K}_2(t)N^2(t)dt}{e^{\int \mathcal{K}_3(t)}dt}+c)}}{e^{-2\int \mathcal{K}_3(t)}dt}\Big[\frac{dr^2}{\chi^2}+r^2(d\theta^2+sin^2\theta d\phi^2)\Big],
  \end{align*} 
     lead to a statistical structure on FLRW space-times.
  	\end{theorem}
Moving on we now study the condition $\pounds_{\xi^i}K^\mu_{\nu\lambda}=0$, for $i=7, 8, 9$. Using (\ref{K9}) and (\ref{AM2}) we get \begin{align*}
&\pounds_{\xi^7}\hat\Gamma^r_{\nu\lambda}
=
\begin{pmatrix}
-\frac{r\kappa}{2{\chi}}&0&0&0\\
0&\frac{r^3\kappa{\chi}\sin^2\theta}{2}&0&0\\
0&0&\frac{r^3\kappa{\chi}}{2}&0\\
0&0&0&	0
\end{pmatrix},\ \ \ \ \ \ \pounds_{\xi^7}\hat\Gamma^\phi_{\nu\lambda}
=
\begin{pmatrix}
0&-\frac{r\kappa}{2\chi}&0&0\\
-\frac{r\kappa}{2\chi}&0&0&0\\
0&0&0&0\\
0&0&0&	0
\end{pmatrix},\\
&
\pounds_{\xi^7}\hat\Gamma^t_{\nu\lambda}
=
\begin{pmatrix}
\frac{a(t)\dot a(t)}{\chi N^2(t)}&0&0&0\\
0&\frac{r^2\chi a(t)\dot a(t)\sin^2\theta}{N^2(t)}&0&0\\
0&0&\frac{r^2\chi a(t)\dot a(t)}{N^2(t)}&0\\
0&0&0&	0
\end{pmatrix},\ \ 
\pounds_{\xi^7}\hat\Gamma^\theta_{\nu\lambda}
=
\begin{pmatrix}
0&0&-\frac{r\kappa}{2\chi}&0\\
0&0&0&0\\
-\frac{r\kappa}{2\chi}&0&0&0\\
0&0&0&	0
\end{pmatrix},
\end{align*} 
\begin{align*}
\pounds_{\xi^8}\hat\Gamma^r_{\nu\lambda}
=\begin{pmatrix}
\cos\theta&0&-r\sin\theta&0\\
0&-r^2\chi^2\cos\theta\sin^2\theta&0&0\\
-r\sin\theta&0&-r^2\chi^2\cos\theta&0\\
0&0&0&	0
\end{pmatrix},
\end{align*}

\begin{align*}
	\pounds_{\xi^8}\hat\Gamma^\phi_{\nu\lambda}
	=\begin{pmatrix}
		0&\cos\theta&0&0\\
		\cos\theta&0&-r\sin\theta&0\\
		0&-r\sin\theta&0&0\\
		0&0&0&	0
	\end{pmatrix},	\ \ 
	\pounds_{\xi^8}\hat\Gamma^\theta_{\nu\lambda}
	=\begin{pmatrix}	\frac{\sin\theta}{r\chi^2}&&\cos\theta&0\\	&r\sin^3\theta&0&0\\
	\cos\theta&0&-r\sin\theta&0\\
	0&0&0&	0
	\end{pmatrix},
\end{align*}

\begin{align*}
\pounds_{\xi^8}\hat\Gamma^t_{\nu\lambda}
=\begin{pmatrix}
\frac{2ra(t)\dot a(t)\cos\theta }{\chi^2 N^2(t)}&0&0&0\\
0&\frac{2r^3a(t)\dot a(t)\cos\theta\sin^2\theta}{N^2(t)}&0&0\\
0&0&\frac{2r^3 a(t)\dot a(t)\cos\theta}{N^2(t)} &0\\
0&0&0&	0
\end{pmatrix},
\end{align*}
\begin{align*}
&\pounds_{\xi^9}\hat\Gamma^r_{\nu\lambda}
=
\begin{pmatrix}
\frac{(\kappa^2r^4-\kappa r^2+2)\cos\theta}{2{\chi^4}}&0&-\frac{(\kappa^2r^4-2\kappa r^2+2)r\sin\theta}{2{\chi^2}}&0\\
0&\frac{r^2\cos\theta\sin^2\theta(3\kappa r^2-2)}{2}&0&0\\
-\frac{(\kappa^2r^4-2\kappa r^2+2)r\sin\theta}{2{\chi^2}}&0&\frac{r^2\cos\theta(3\kappa r^2-2)}{2}&0\\
0&0&0&	0
\end{pmatrix},
\end{align*}
\begin{align*}
&\pounds_{\xi^9}\hat\Gamma^\phi_{\nu\lambda}
=
\begin{pmatrix}
0&\cos\theta&0&0\\
\cos\theta&0&-r\sin\theta&0\\
0&-r\sin\theta&0&0\\
0&0&0&	0
\end{pmatrix},
\end{align*}
\begin{align*}
&\pounds_{\xi^9}\hat\Gamma^\theta_{\nu\lambda}
=
\begin{pmatrix}
-\frac{\sin\theta(3\kappa r^2-2)}{2r\chi^2}&0&\cos\theta&0\\
0&-\frac{r\sin^3\theta(\kappa r^2-2)}{2}&0&0\\
\cos\theta&0&-\frac{r\sin\theta(\kappa r^2+2)}{2}&0\\
0&0&0&	0
\end{pmatrix},
\end{align*}
\begin{align*}
&\pounds_{\xi^9}\hat\Gamma^t_{\nu\lambda}
=
\begin{pmatrix}
-\frac{ra(t)\dot a(t)\cos\theta(\kappa r^2-2)}{\chi^4 N^2(t)}&0&-\frac{r^4\kappa a(t)\dot a(t)\sin\theta}{2\chi^2 N^(t)}&0\\
0&\frac{2r^3a(t)\dot a(t)\cos\theta\sin^2\theta}{N^2(t)}&0&0\\
-\frac{r^4\kappa a(t)\dot a(t)\sin\theta}{2\chi^2 N^2(t)}&0&\frac{2r^3a(t)\dot a(t)\cos\theta}{N^2(t)}&0\\
0&0&0&	0
\end{pmatrix},
\end{align*}
Now we must obtain $\pounds_{\xi^i}\Gamma^\mu_{\kappa\lambda}$, $i=7, 8, 9$ for statistical connections determined by (\ref{9'}). Since we need the Lie derivative of connections (which are neither metric-compatible, nor torsion-free) with respect to $\xi^i$, $i=7, 8, 9$  in the following, we calculate $\pounds_{\xi^i}\Gamma^\mu_{\kappa\lambda}$, $i=7, 8, 9$ for connections given by (\ref{9}). Using (\ref{9}) we get
\begin{align*}
&\pounds_{\xi^7}\Gamma^r_{\nu\lambda}
=
\begin{pmatrix}
-\frac{r\kappa}{2{\chi}}&0&0&0\\
0&\frac{r^3\kappa{\chi}\sin^2\theta}{2}&\frac{r^2\chi^2\sin\theta\mathcal{K}_5(t)}{2}&0\\
0&-\frac{r^2\chi^2\sin\theta\mathcal{K}_5(t)}{2}&\frac{r^3\kappa{\chi}}{2}&0\\
0&0&0&	0
\end{pmatrix},\ \ \ \ \ \pounds_{\xi^7}\Gamma^\phi_{\nu\lambda}
=
\begin{pmatrix}
0&-\frac{r\kappa}{2\chi}&-\frac{\mathcal{K}_5(t)}{2\sin\theta}&0\\
-\frac{r\kappa}{2\chi}&0&0&0\\
\frac{\mathcal{K}_5(t)}{2\sin\theta}&0&0&0\\
0&0&0&	0
\end{pmatrix},\\
&
\pounds_{\xi^7}\Gamma^t_{\nu\lambda}
=
\begin{pmatrix}
\frac{\mathcal{K}_2(t)}{\chi}&0&0&0\\
0&r^2\chi\sin^2\theta\mathcal{K}_2(t)&0&0\\
0&0&r^2\chi\mathcal{K}_2(t)&0\\
0&0&0&	0
\end{pmatrix},\ \ \ \ \ 
\pounds_{\xi^7}\Gamma^\theta_{\nu\lambda}
=
\begin{pmatrix}
0&\frac{\sin\theta\mathcal{K}_5(t)}{2}&-\frac{r\kappa}{2\chi}&0\\
-\frac{\sin\theta\mathcal{K}_5(t)}{2}&0&0&0\\
-\frac{r\kappa}{2\chi}&0&0&0\\
0&0&0&	0
\end{pmatrix},
\end{align*} 
\begin{align*}
\pounds_{\xi^8}\Gamma^r_{\nu\lambda}
=\begin{pmatrix}
\cos\theta&0&-r\sin\theta&0\\
0&-r^2\chi^2\cos\theta\sin^2\theta&\frac{r^3\chi\sin2\theta K_5(t)}{2}&0\\
-r\sin\theta&-\frac{r^3\chi\sin2\theta K_5(t)}{2}&-r^2\chi^2\cos\theta&0\\
0&0&0&	0
\end{pmatrix},
\end{align*} 
\begin{align*}
\pounds_{\xi^8}\Gamma^\phi_{\nu\lambda}
=\begin{pmatrix}
0&\cos\theta&-\frac{r\cot\theta \mathcal{K}_5(t)}{\chi}&0\\
\cos\theta&0&-r\sin\theta&0\\
\frac{r\cot\theta \mathcal{K}_5(t)}{\chi}&-r\sin\theta&0&0\\
0&0&0&	0
\end{pmatrix},
\ \ 
\pounds_{\xi^8}\Gamma^\theta_{\nu\lambda}
=\begin{pmatrix}
\frac{\sin\theta}{r\chi^2}&\frac{r\sin2\theta \mathcal{K}_5(t)}{2\chi}&\cos\theta&0\\
-\frac{r\sin2\theta \mathcal{K}_5(t)}{2\chi}&r\sin^3\theta&0&0\\
\cos\theta&0&-r\sin\theta&0\\
0&0&0&	0
\end{pmatrix},
\end{align*} 
\begin{align*}
\pounds_{\xi^8}\Gamma^t_{\nu\lambda}
=\begin{pmatrix}
\frac{2r\cos\theta \mathcal{K}_2(t)}{\chi^2}&0&0&0\\
0&2r^3\cos\theta\sin^2\theta \mathcal{K}_2(t)&0&0\\
0&0&2r^3\cos\theta \mathcal{K}_2(t)&0\\
0&0&0&	0
\end{pmatrix},
\end{align*}
\begin{align*}
&\pounds_{\xi^9}\Gamma^r_{\nu\lambda}
=
\begin{pmatrix}
\frac{(\kappa^2r^4-\kappa r^2+2)\cos\theta}{2{\chi^4}}&\frac{r^4\kappa\sin^2\theta\mathcal{K}_5(t)}{2\chi}&-\frac{(\kappa^2r^4-2\kappa r^2+2)r\sin\theta}{2{\chi^2}}&0\\
-\frac{r^4\kappa\sin^2\theta\mathcal{K}_5(t)}{2\chi}&\frac{r^2\cos\theta\sin^2\theta(3\kappa r^2-2)}{2}&-\frac{r^3\sin2\theta(3\kappa r^2-2)\mathcal{K}_5(t)}{4\chi}&0\\
-\frac{(\kappa^2r^4-2\kappa r^2+2)r\sin\theta}{2{\chi^2}}&\frac{r^3\sin2\theta(3\kappa r^2-2)\mathcal{K}_5(t)}{4\chi}&\frac{r^2\cos\theta(3\kappa r^2-2)}{2}&0\\
0&0&0&	0
\end{pmatrix},
\end{align*}
\begin{align*}
&\pounds_{\xi^9}\Gamma^\phi_{\nu\lambda}
=
\begin{pmatrix}
0&\cos\theta&\frac{r\cot\theta(\kappa r^2-2)\mathcal{K}_5(t)}{2\chi^3}&0\\
\cos\theta&0&-r\sin\theta&0\\
-\frac{r\cot\theta(\kappa r^2-2)\mathcal{K}_5(t)}{2\chi^3}&-r\sin\theta&0&0\\
0&0&0&	0
\end{pmatrix},
\end{align*}
\begin{align*}
&\pounds_{\xi^9}\Gamma^\theta_{\nu\lambda}
=
\begin{pmatrix}
-\frac{\sin\theta(3\kappa r^2-2)}{2r\chi^2}&-\frac{r\sin2\theta(\kappa r^2-2)\mathcal{K}_5(t)}{4\chi^3}&\cos\theta&0\\
\frac{r\sin2\theta(\kappa r^2-2)\mathcal{K}_5(t)}{4\chi^3}&-\frac{r\sin^3\theta(\kappa r^2-2)}{2}&\frac{r^4\kappa\sin^2\theta\mathcal{K}_5(t)}{2\chi}&0\\
\cos\theta&-\frac{r^4\kappa\sin^2\theta\mathcal{K}_5(t)}{2\chi}&-\frac{r\sin\theta(\kappa r^2+2)}{2}&0\\
0&0&0&	0
\end{pmatrix},
\end{align*}
\begin{align*}
&\pounds_{\xi^9}\Gamma^t_{\nu\lambda}
=
\begin{pmatrix}
-\frac{r\cos\theta(\kappa r^2-2)\mathcal{K}_2(t)}{\chi^4}&0&-\frac{r^4\kappa\sin\theta\mathcal{K}_2(t)}{2\chi^2}&0\\
0&2r^3\cos\theta\sin^2\theta\mathcal{K}_2(t)&0&0\\
-\frac{r^4\kappa\sin\theta\mathcal{K}_2(t)}{2\chi^2}&0&2r^3\cos\theta\mathcal{K}_2(t)&0\\
0&0&0&	0
\end{pmatrix},
\end{align*}
Setting $\mathcal{K}_5=0$  and using the form (\ref{1000}) in the above matrices, we have $\pounds_{\xi^i}\Gamma^\mu_{\kappa\lambda}$, $i=7, 8, 9$ for statistical connections determined by (\ref{9'}). Therefore we get
\begin{align*}
\pounds_{\xi^7}K^t_{\nu\lambda}
=
\begin{pmatrix}
\frac{\mathcal{A}(t)}{\chi}&0&0&0\\
0&r^2\chi\sin^2\theta\mathcal{A}(t)&0&0\\
0&0&0&r^2\chi\mathcal{A}(t)\\
0&0&0&	0
\end{pmatrix},
\end{align*}
\begin{align*}
\pounds_{\xi^8}K^t_{\nu\lambda}
=\begin{pmatrix}
\frac{2r\cos\theta \mathcal{A}(t)}{\chi^2}&0&0&0\\
0&2r^3\cos\theta\sin^2\theta \mathcal{A}(t)&0&0\\
\cos\theta&0&2r^3\cos\theta \mathcal{A}(t)&0\\
0&0&0&	0
\end{pmatrix},
\end{align*}
\begin{align*}
&\pounds_{\xi^9}K^t_{\nu\lambda}
=
\begin{pmatrix}
-\frac{r\cos\theta(\kappa r^2-2)\mathcal{A}(t)}{\chi^4}&0&-\frac{r^4\kappa\sin\theta\mathcal{A}(t)}{2\chi^2}&0\\
0&2r^3\cos\theta\sin^2\theta\mathcal{A}(t)&0&0\\
-\frac{r^4\kappa\sin\theta\mathcal{A}(t)}{2\chi^2}&0&2r^3\cos\theta\mathcal{A}(t)&0\\
0&0&0&	0
\end{pmatrix},
\end{align*}
where $\mathcal{A}(t)=\mathcal{K}_2(t)N^2(t)-a(t)\dot a(t)$. So, we get
$\pounds_{\xi^i}K^\mu_{\nu\lambda}=0$, $i=7, 8, 9$ if and only if  $\mathcal{A}(t)=0$. Setting (\ref{LLLLLL}) in (\ref{10000}), it follows 
\[2\dot a(t)a(t)-\mathcal{K}_3(t)a^2(t)-\mathcal{K}_2(t)N^2(t)=0,\]
Putting $a(t)\dot a(t)=\mathcal{K}_2(t)N^2(t)$ in the last equation,
we deduce $\dot a(t)a(t)-\mathcal{K}_3(t)a^2(t)=0$, which gives
\[
a(t)=c\ e^{\int \mathcal{K}_3(t)dt}.
\]
\begin{theorem}
The vector fields $\xi^i$, $i=1, \cdots, 9$ given by Corollary \ref{AM6}, are CKVF's for FLRW space-times equipped with statistical connections obtained by (\ref{9'}) and  FLRW metric  determined by
 \begin{align*}
g=-N^2(t)dt^2+(c\ e^{\int \mathcal{K}_3(t)dt})^2\Big[\frac{dr^2}{\chi^2}+r^2(d\theta^2+sin^2\theta d\phi^2)\Big].
\end{align*} 
\end{theorem}
As we mentioned in the above, the connections given by (\ref{9}) are all connections satisfying $\pounds_{\xi^i}\Gamma^\mu_{\kappa\lambda}=0$, for $i=1, \cdots, 6$. Now the question that arises is  whether these connections satisfy also  $\pounds_{\xi^i}\Gamma^\mu_{\nu\lambda}=0$, for $i=7, 8, 9$ or under what conditions do these connections satisfy $\pounds_{\xi^i}\Gamma^\mu_{\nu\lambda}=0$, for $i=7, 8, 9$. According to the matrix forms of $\pounds_{\xi^8}\Gamma^\mu_{\nu\lambda}$ and $\pounds_{\xi^9}\Gamma^\mu_{\nu\lambda}$, it clearly follows that these cannot be zero. Moreover, the matrix form of $\pounds_{\xi^7}\Gamma^\mu_{\nu\lambda}$ shows that $\pounds_{\xi^7}\Gamma^\mu_{\nu\lambda}=0$ if and only if $\kappa=0$. Therefore we conclude the following: 
\begin{theorem}
	When $\kappa\neq 0$, there is no  proper CKVF $X$ satisfying $\pounds_X\Gamma^\mu_{\nu\lambda}=0$. Moreover, in the case $\kappa=0$, the vector field generated by $\{\xi^1, \cdots, \xi^7\}$ is a CKVF that satisfies $\pounds_X\Gamma^\mu_{\nu\lambda}=0$.
\end{theorem}
\subsection{Lie derivative of the torsion tensor} In this part, we focus on the connections given by (\ref{9}).
In local coordinates (i.e., in a holonomic frame) the torsion $T$
is given by
\begin{align}
&T^\lambda_{\mu\nu}=\Gamma^\lambda_{\mu\nu}-\Gamma^\lambda_{\nu\mu}.
\end{align}
When $\Gamma^\lambda_{\mu\nu}$ is symmetric with respect to $\mu$ and $\nu$, then we have $T^\lambda_{\mu\nu}=0$ and we say that the connection is torsion-free (or symmetric). Torsion-free connections are important in mathematics and physics. The famous Levi-Civita connection is a torsion-free connection. Also, statistical connections as the main tools of information geometry are torsion-free. On the other hand, non-symmetric connections (connections with non-zero torsion) have also many applications in mathematics and physics. Indeed, the use of non-symmetric connections became especially relevant after the appearance of the works of Einstein, related to creating a possible Unified Field Theory (UFT). It is also known that fermionic matter induces space torsion (see for instance \cite{Hehl:1976kj}). This is the case in the so-called Einstein-Cartan\footnote{There also exist the teleparallel gravity theories where the connection is flat and metric compatible and and gravity is mediated through torsion (see for instance \cite{Aldrovandi:2013wha}).} theory of gravity which is a viable extention of GR that incorporates also the spin of matter \cite{Hehl:1976kj}. Torsion also appears in condensed matter systems and in particular in the theory of dislocations. From the above, it becomes clear that torsion plays a significant role in physics. It then stands to reason to study its mathematical properties in depth.

Let us focus on the Lie derivative here. Obviously, the Lie derivative of the torsion tensor $T^\lambda_{\mu\nu}$ satisfies the following
\begin{align*}
\pounds_XT^\mu_{\nu\lambda}=\pounds_X\Gamma^\mu_{\nu\lambda}-\pounds_X\Gamma^\mu_{\lambda\nu}.
\end{align*}
The torsion $T$	is symmetric under a group action if $\pounds_XT^\mu_{\nu\lambda}=0$,  for all generating vector fields $X$. Since $\pounds_{\xi^i}\Gamma^\mu_{\nu\lambda}=0$, for $i=1, \cdots 6$, then $\pounds_{\xi^i}T^\mu_{\nu\lambda}=0$. Indeed, the Lie derivative of torsion tensor with respect to Killing vector fields is zero. Now we study it for proper CKVF's by computing $\pounds_{\xi^i}T^\mu_{\nu\lambda}$, $i=7, 8, 9$.

From the above matrices, it follows that
\begin{align*}
&\pounds_{\xi^7}T^r_{\phi\theta}
=-\pounds_{\xi^7}T^r_{\theta\phi}=r^2(\chi^2)\mathcal{K}_5(t)\sin\theta,\\
&\pounds_{\xi^7}T^\phi_{r\theta}
=-\pounds_{\xi^7}T^\phi_{\theta r}=-\frac{\mathcal{K}_5(t)}{\sin\theta},\\
&\pounds_{\xi^7}T^\theta_{r\phi}
=-\pounds_{\xi^7}T^\theta _{\phi r}={\mathcal{K}_5(t)}{\sin\theta}.
\end{align*} 
Similarly, we get  
\begin{align*}
&\pounds_{\xi^8}T^r_{\phi\theta}
=-\pounds_{\xi^8}T^r_{\theta\phi}=r^3\chi\mathcal{K}_5(t)\sin2\theta, \ \ \ \ \ \ \ \ \ \ \ \ \ \ \ \ \ \pounds_{\xi^9}T^\theta_{r\phi}=-\pounds_{\xi^9}T^\theta_{ \phi r }=-\frac{ r(\kappa r^2-2)\mathcal{K}_5(t)\sin2\theta}{2({\chi^2})^{\frac{3}{2}}},\\
&\pounds_{\xi^8}T^\phi_{r\theta}
=-\pounds_{\xi^8}T^\phi_{\theta r}=-\frac{2r\mathcal{K}_5(t)\cot\theta}{\chi}, \ \ \ \ \ \ \ \  \ \ \ \ \ \ \ \ \ \pounds_{\xi^9}T^r_{\phi\theta}=-\pounds_{\xi^9}T^r_{\theta\phi }=-\frac{ r^3(3\kappa r^2-2)\mathcal{K}_5(t)\sin2\theta}{2\chi},\\
&\pounds_{\xi^8}T^\theta_{r\phi}
=-\pounds_{\xi^8}T^\theta _{\phi r}=\frac{r\mathcal{K}_5(t)\sin2\theta}{\chi},\ \ \ \ \ \ \ \  \ \ \ \ \ \ \  \ \ \ \ \pounds_{\xi^9}T^\phi_{r\theta}=-\pounds_{\xi^9}T^\phi_{\theta r }=\frac{ r(\kappa r^2-2)\mathcal{K}_5(t)\cot\theta}{({\chi^2})^{\frac{3}{2}}},\\
& \ \ \ \ \ \ \ \  \ \ \ \ \ \ \  \ \ \ \ \ \ 
\ \ \  \pounds_{\xi^9}T^r_{r\phi}=-\pounds_{\xi^9}T^r_{\phi r}=\frac{\kappa r^4\mathcal{K}_5(t)\sin^2\theta}{\chi}=\pounds_{\xi^9}T^\theta_{\phi\theta}
=-\pounds_{\xi^9}T^\theta _{ \theta\phi},
\end{align*} 
In conclusion we have the following.
\begin{proposition}
	On the FLRW space-time with the connection given by (\ref{9}), the equation $\pounds_{\xi^i}T^\mu_{\nu\lambda}=0, i=1,\cdots,9$ is valid if and only if $\mathcal{K}_5(t)=0$.
\end{proposition}
\section{Conclusion}
We have revisited the probelm of determining conformal Killing vector fields for the FLRW spacetime. Our method was based simply on solving partial differential equations (new point of view). Furthermore,  we initiated the study of  statistical structures on  FLRW spacetimes and obtained a classification of statistical structures  admiting such conformal vector fields. In brief, we studied conformal Killing vector fields  from a statistical manifold point of view.\\
Moreover, we have also considered torsionful connections and have investigated the symmetries of the torsion tensor. For the latter, a complete list of conformal vector feilds in FLRW spacetimes was derived.\\
In future works, many cases can be studied on these space-times from a statistical manifold point of view. For example, one could conduct an investigation of the statistical Ricci-soliton equations along with  2-Killing vector fields. It is also possible to study the tangent bundle created by FLRW spacetime along with the corresponding related concepts such as curvature tensor, Ricci tensor, sectional curvature and sectional $K$-curvature.

\end{document}